\documentclass[12pt]{article}
\usepackage{times}
\usepackage{titling}
\usepackage{graphicx}
\usepackage{amsmath}
\usepackage{color}
\usepackage{bm}

   \definecolor{darkgreen}{RGB}{0,100,0}

        \title{\bf Fractional excitations in the square lattice quantum antiferromagnet} 

        \author{B. Dalla Piazza,$^{1,\ast}$ M. Mourigal,$^{1,2,3,\ast}$ N. B. Christensen,$^{4,5}$ \\ 
            G. J. Nilsen,$^{1,6}$ P. Tregenna-Piggott,$^{5}$ T. G. Perring,$^7$ \\
            M. Enderle,$^2$ D. F. McMorrow,$^8$ D. A. Ivanov,$^{9,10}$ and H. M. R\o{}nnow$^1$\\
            \\
            \normalsize{$^{1}$Laboratory for Quantum Magnetism,}\\
            \normalsize{\'Ecole Polytechnique F\'ed\'erale de Lausanne (EPFL), CH-1015, Switzerland}\\
            \normalsize{$^2$Institut Laue-Langevin, BP 156,}\\
            \normalsize{F-38042 Grenoble Cedex 9, France}\\
            \normalsize{$^3$Institute for Quantum Matter and Department of Physics and Astronomy,}\\
            \normalsize{Johns Hopkins University, Baltimore, MD 21218, USA}\\
            \normalsize{$^4$Department of Physics, Technical University of Denmark (DTU), DK-2800 Kgs. Lyngby, Denmark}\\
            \normalsize{$^5$Laboratory for Neutron Scattering, Paul Scherrer Institute,}\\
            \normalsize{CH-5232 Villigen PSI, Switzerland}\\
            \normalsize{$^6$Department of Chemistry, University of Edinburgh,}\\
            \normalsize{Edinburgh, EH9 3JJ, United Kingdom}\\
            \normalsize{$^7$ISIS Facility, Rutherford Appleton Laboratory,}\\
            \normalsize{Chilton, Didcot, Oxon OX11 OQX, United Kingdom,}\\
            \normalsize{$^8$London Centre for Nanotechnology and Department of Physics and Astronomy,}\\
            \normalsize{University College London, London WC1H 0AH, United Kingdom}\\
            \normalsize{$^9$Institute for Theoretical Physics, ETH Z\"urich,}\\
            \normalsize{CH-8093 Z\"urich, Switzerland}\\
            \normalsize{$^{10}$Institute for Theoretical Physics, University of Z\"urich,}\\ 
            \normalsize{CH-8057 Z\"urich, Switzerland}\\
            \\
            \normalsize{$^\ast$To whom correspondence should be addressed; E-mail: bastien.dallapiazza@epfl.ch, mourigal@jhu.edu}
        }

        \date{}

\begin{document} 
        \baselineskip24pt
        \maketitle

        \begin{abstract}
            The square-lattice quantum Heisenberg antiferromagnet displays a pronounced anomaly of unknown origin in its magnetic excitation spectrum. The anomaly manifests itself only for short wavelength excitations propagating along the direction connecting nearest neighbors. Using polarized neutron spectroscopy, we have fully characterized the magnetic fluctuations in the model metal-organic compound CFTD, revealing an isotropic continuum at the anomaly indicative of fractional excitations. A theoretical framework based on the Gutzwiller projection method is developed to explain the origin of the continuum at the anomaly. This indicates that the anomaly arises from deconfined fractional spin-1/2 quasiparticle pairs, the 2D analog of 1D spinons. Away from the anomaly the conventional spin-wave spectrum is recovered as pairs of fractional quasiparticles bind to form spin-1 magnons. Our results therefore establish the existence of fractional quasiparticles in the simplest model two dimensional antiferromagnet even in the absence of frustration.
        \end{abstract}

        A fascinating manifestation of quantum mechanics in the presence of many-body interactions is the emergence of elementary excitations carrying fractional quantum numbers. 
Fractional excitations were a central ingredient to understand the anomalous quantum Hall effect~\cite{laughlin_1999}, and have been investigated in a range of systems including conducting polymers~\cite{su_1979}, bilayer graphene~\cite{hou_2007}, cold atomic gases~\cite{simon_2011}, and low-dimensional quantum magnets~\cite{baskaran_1987,balents_2010}. Among the latter class of systems, the spin-1/2 chain quantum Heisenberg antiferromagnet (HAF) is perhaps the simplest realization of this paradigm, for which the ground-state and the excitation spectrum known exactly \cite{bethe_1931,fadeev_1981,Mueller_1981}. The elementary excitations of an antiferromagnet in the large-spin, semi-classical case are well-defined spin-waves (magnons) created by $\Delta S=1$ processes, while in the quantum  spin-1/2 chain a $\Delta S=1$ process generate pairs of unbound fractional quasi-particles known as spinons each carrying a $S=1/2$ quantum number~\cite{fadeev_1981}. While the existence of spinons in the spin-1/2 HAF chain has been established in a number of experimental studies on model 1D materials~\cite{tennant_1995,lake_2005,mourigal_2013}, establishing their existence in higher dimensions is an ongoing challenge\cite{balents_2010}. To date the main candidate systems comprise antiferromagnets with frustrated geometries, including triangular \cite{coldea_2001} or kagome \cite{han_2012,jeong_field-induced_2011,kozlenko_2012} lattices. In this work, we take a frustration-free route and focus on the spin-1/2 square-lattice quantum Heisenberg antiferromagnet (SLQHAF), one of the most fundamental models in magnetism. It is defined by the Hamiltonian
        \begin{equation} \label{heis_spin}
            {\cal H}=J\sum_{\langle i,j \rangle}\mathbf{S}_i\cdot \mathbf{S}_j \ , 
        \end{equation} 
        where the antiferromagnetic exchange $J$ is restricted to nearest-neighbor spins ${\langle i,j\rangle}$. We provide experimental and theoretical evidence that even in this simplest of 2D models deconfined fractional $S=1/2$ quasiparticles exist and are responsible for the very significant deviations of the high-energy, short-wavelength spin dynamics from conventional spin-wave theory.

        It may seem surprising that the SLQHAF is a candidate for hosting fractional excitations as at a superficial level its magnetic order and excitations resemble those of a classical system. It is well established that at zero temperature quantum fluctuations are insufficient to prevent the development of collinear antiferromagnetic order~\cite{manousakis_1991}. The elementary excitations of this N\'eel state, when calculated using semi-classical spin wave theory (SWT), can be described as $S=1$ bosonic quasiparticles, known as magnons: the one-magnon dispersion is gapless~\cite{anderson_1952,kubo_1952}, with two-magnon excitations occupying a continuum at higher energy. The interaction between magnons is relatively weak and leads to an upward renormalization of the magnon energy~\cite{dyson_1956,ogushi_1960} and to scattering between two-magnon states~\cite{harris_1971,canali_1993}. 

        While none of the above properties suggest the existence of quasiparticle fractionalization, quantum effects are nevertheless far from negligible in the SLQHAF. This is evidenced by the observation that quantum zero-point fluctuations reduce the staggered moment to only 62\% of its fully ordered value $S\!=\! 1/2$~\cite{reger_1988,hamer_1992}. This suggests that the SLQHAF may in fact be close to a state lacking spin-symmetry breaking, such as the resonating-valence-bond (RVB) state proposed for the cuprate realization of this model by Anderson \cite{anderson_1987}. In particular, fractional spin excitations present in the RVB state may be relevant for the spin dynamics in the N\'eel state, especially at high energies. Indeed, analytical theories using bosonic~\cite{auerbach_1988} or fermionic~\cite{hsu_1990,ho_2001} fractional quasiparticles have long been proposed. By analogy with the 1D case, these are referred to as spinons.

        \begin{figure}[t!]
            \centering
            \includegraphics[width=\linewidth]{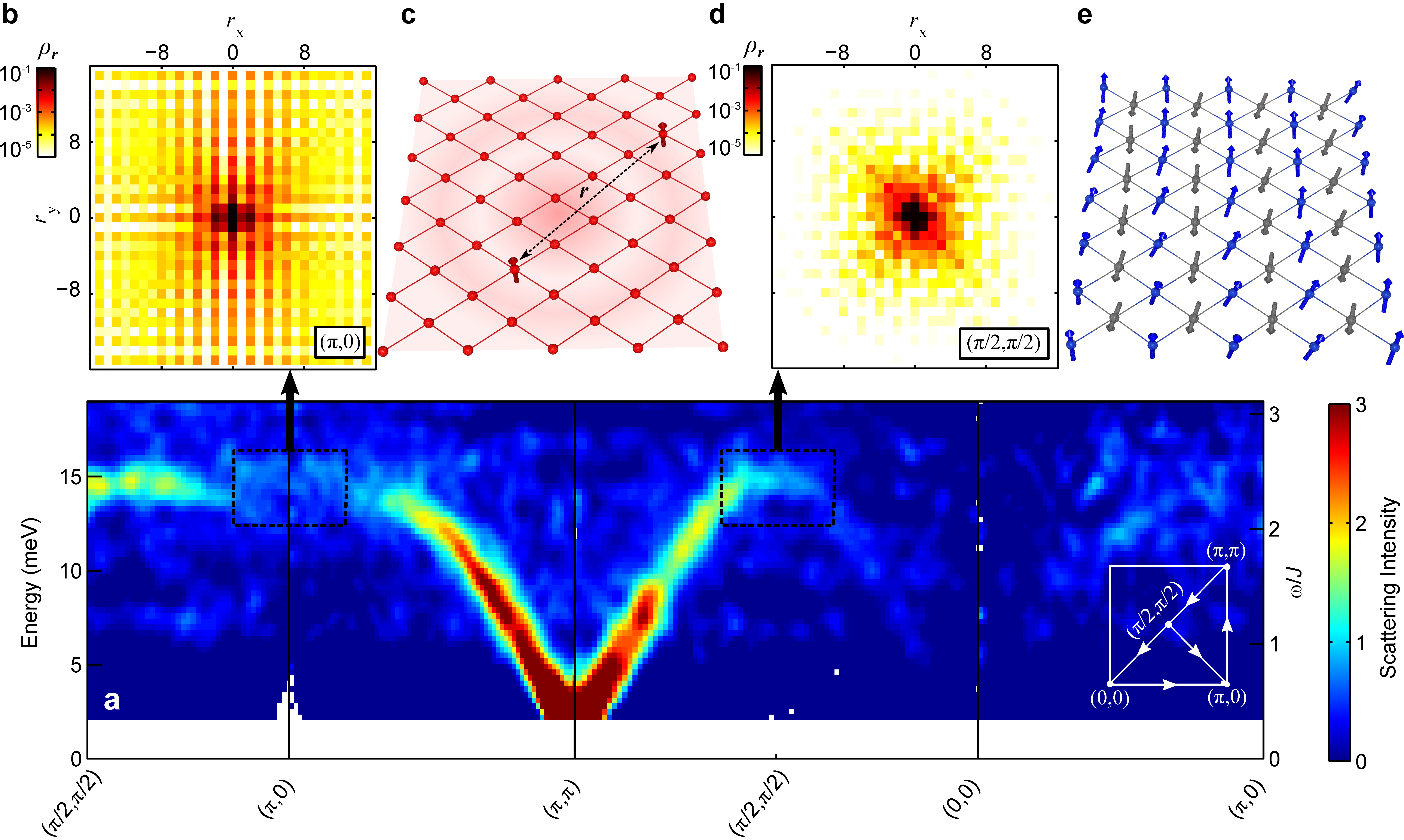}
            \caption{\label{fig1}  ({\bf a}) Overview of the magnetic excitation spectrum of CFTD. Momentum and energy dependence of the (total) dynamic structure factor ${\cal S}({\bf q},\omega)$ measured by time-of-flight inelastic neutron scattering. Square boxes (black dashed) highlight the $(\pi,0)$ and $(\pi/2,\pi/2)$ wave-vectors. ({\bf b}) and ({\bf d}) Corresponding distributions of real-space fractional quasiparticle-pair separations, as calculated in the $|\mbox{SF}\rangle$ variational state (Eq.\ \ref{rho}), evidencing respectively the unbound and bound nature of the pair excitations. ({\bf c}) Pictorial representation of a quasiparticle pair excitation. ({\bf e}) Pictorial representation of a spin-wave excitation (magnon).}
        \end{figure}

        The magnetic excitation spectrum of various realizations of the SLQHAF have been investigated using neutron spectroscopy, including parent compounds of the high-$T_c$ cuprates such as La$_2$CuO$_4$~\cite{coldea_spin_2001,headings_2010} and the metal-organic compound Cu(DCOO)$_2\cdot$4D$_2$O (CFTD) \cite{ronnow_2001,christensen_2007} considered here. These experiments have established that while SWT gives an excellent account of the low-energy spectrum, a glaring anomaly is present at high energy for wavevectors in the vicinity of $(\pi,0)$, where $\mathbf{q}=(q_x,q_y)$ is expressed in the square-lattice Brillouin-zone of unit-length $2\pi$. The anomaly is evident as a dramatic wipeout of intensity of the otherwise sharp excitations, Fig. \ref{fig1}{\bf a} for CFTD and Ref.~\cite{headings_2010} for La$_2$CuO$_4$, and as a $7\%$ downward dispersion along the magnetic zone-boundary connecting the $(\pi/2,\pi/2)$ and $(\pi,0)$ wavevectors for CFTD (see also Fig. \ref{fig4}{\bf d}). Unambiguously identifying the origin of this effect is complicated by the presence, in some materials, of small additional exchange terms such as electronic ring-exchange or further neighbor exchange. In contrast, the deviations observed in CFTD agree with numerical results obtained by series expansion \cite{singh_1995,zheng_2005}, quantum Monte Carlo \cite{syljuasen_2000,sandvik_2001}, and exact diagonalization \cite{luscher_2009} methods for the model of Eq.~(\ref{heis_spin}), proving that the anomaly is in this case {\it intrinsic}~\cite{christensen_2007}. Given the fundamental nature of the SLQHAF, it is clearly desirable to develop a microscopic understanding of the origin of the anomaly.

        Here we present polarized neutron scattering results on CFTD which establish the existence of a \emph{spin-isotropic} continuum at $(\pi,0)$, which contrasts sharply with the dominantly longitudinal continuum at $(\pi/2,\pi/2)$ and with expectations for a magnetically ordered ground state. Using a fermionic description of the spin dynamics based on a Gutzwiller-projected variational approach, we argue that this is a signature of {\it unbound} pairs of fractional $S\!=\!1/2$ quasiparticles (Fig.~\ref{fig1}{\bf b} and \ref{fig1}{\bf c}) at the $(\pi,0)$ wave vector. At other wave vectors, including $(\pi/2,\pi/2)$ (Fig.~\ref{fig1}{\bf d}), our approach yields {\it bound}  pairs these fractional quasiparticles and so recovers a conventional magnon spectrum, in agreement with SWT (Fig.~\ref{fig1}{\bf e}).

        \begin{figure}[t!]
            \centering
            \includegraphics[width=0.75\linewidth]{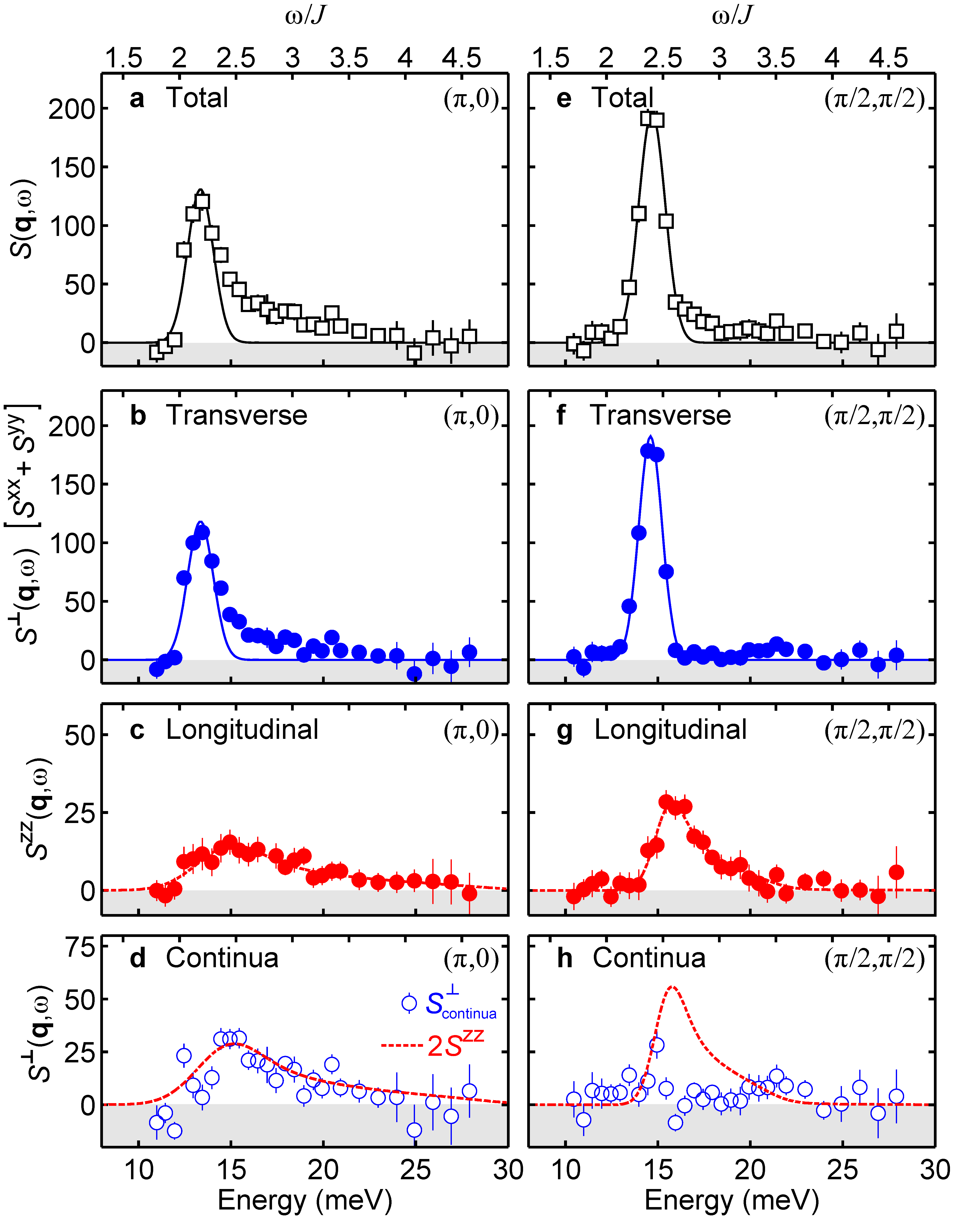}
            \caption{\label{fig2} Summary of the polarized neutron scattering data.
            Energy dependence of the total, transverse and longitudinal contributions to the dynamic structure factor at constant wave vectors ({\bf a}--{\bf d}) ${\bf q}=(\pi,0)$ and ({\bf e}--{\bf h}) ${\bf q}=(\pi/2,\pi/2)$ measured by polarized neutron scattering on CFTD. The solid lines indicate resolution-limited Gaussian fits, while the dashed lines are empirical lineshapes used as guides-to-the-eye.}
        \end{figure}

        Neutron scattering experiments were performed on single crystals of CFTD using unpolarized time-of-flight spectroscopy (Fig.~\ref{fig1}) and triple-axis spectroscopy with longitudinal polarization analysis (see Supplementary Materials). The results of our polarized experiment are presented in Fig.~\ref{fig2} through the energy dependence of the diagonal components of the dynamic structure factor. By combining wave-vectors from different equivalent Brillouin zones (see Supplementary Materials), we can reconstruct the total dynamic structure factor (Fig.~\ref{fig2}{\bf a} and {\bf e}) and separate contributions from spin fluctuations that are transverse to and along (Fig.~\ref{fig2}{\bf b}-{\bf c} and {\bf f}-{\bf e}) the ordered moment. Within SWT, the resulting transverse and longitudinal spectra are dominated by one-magnon and two-magnon excitations, respectively. At $(\pi/2,\pi/2)$ and at an energy of $\omega/J\!=\!2.38(2)$ we observe a sharp, energy resolution limited  peak ($\Delta\omega\!=\!1.47(5)$~meV $=\!0.24(1)\,J$, FWHM) which is the signature of a long-lived, single-particle excitation (Fig.~\ref{fig2}{\bf e}). Most of the observed spectral weight is in the resolution-limited peak of the transverse channel $\mathcal{S}^\perp(\mathbf{q},\omega)\equiv\mathcal{S}^{xx}(\mathbf q,\omega)+\mathcal{S}^{yy}(\mathbf q,\omega)$ (Fig.~\ref{fig2}{\bf f}), while a weak continuum extends from $\omega/J\!\approx\!2.3$ to $3.4$ with a maximum around $\omega/J\!\approx\!2.6$ in the longitudinal channel, $\mathcal{S}^{zz}(\mathbf q,\omega)$ (Fig.~\ref{fig2}{\bf g}). In contrast, the response at $(\pi,0)$ displays a pronounced high-energy tail, starting right above the peak maximum at $\omega/J=2.19(2)$ and extending up to $\omega/J\!\approx\!3.8$. This tail carries 40(12)\% of the total spectral weight at $(\pi,0)$ (Fig.~\ref{fig2}{\bf a}) and is evident in both the transverse (Fig.~\ref{fig2}{\bf b}) and longitudinal (Fig.~\ref{fig2}{\bf c}) channels. To isolate the continuous component in the transverse channel we subtract resolution-limited Gaussians corresponding to sharp, single-particle responses, with the results shown in Fig.~\ref{fig2}{\bf d} and {\bf h}. This analysis reveals the important fact that the transverse continuum at $(\pi,0)$ is within error  {\it twice} the longitudinal contribution (Fig.~\ref{fig2}{\bf d}). Thus we can conclude that the continuum at $(\pi,0)$ arises from correlations which are isotropic in spin space with ${\cal S}^{\perp}(\mathbf q,\omega)=2{\cal S}^{zz}({\mathbf q},\omega)$, while by contrast the continuum contribution at $(\pi/2,\pi/2)$ is fully contained in the longitudinal channel (Fig.~\ref{fig2}{\bf h}).

        The pronounced asymmetric and non-Lorentzian lineshape of the continuum at $(\pi,0)$ cannot be accounted for by conventional effects, even including instrumental resolution. SWT predicts that magnon interaction transfer up to 20\% of the transverse spectral weight at the zone-boundary from the sharp one-magnon peak to a higher energy continuum of three-magnon states~\cite{canali_1993}. However, the resulting lineshape differs radically from our observations, does not coincide with the longitudinal response, and does not appear to depend significantly on the wave vector along the zone boundary. Spontaneous magnon decays can in principle produce an asymmetric lineshape but are prohibited in this case by the collinearity of the magnetic order~\cite{harris_1971,zhitomirsky_2013}. Instead, recent quantum Monte Carlo work~\cite{sandvik_2013} suggests to look for explanations of the continuum contribution to the dynamic structure factor at $(\pi,0)$ involving the deconfinement of fractional excitations. This is further motivated by the observed coexistence of sharp two-spinon bound states with a broad multi-spinon continuum, at comparable energy ranges but different wave-vectors, in the quasi-2D materials Cs$_2$CuCl$_4$~\cite{coldea_2001,kohno_2007} and LiCuVO$_4$~\cite{enderle_2010} made of strongly-coupled Heisenberg chains.

        In order to explore whether fractionalization of magnons can account for the $(\pi,0)$ anomaly in the SLQHAF, we use a theoretical approach based on Gutzwiller-projected variational wave functions\cite{gros_1988,gros_1989}. In this approach, spin operators are transformed into \emph{pairs} of $S\!=\!1/2$ fermionic operators so that Eq.~\ref{heis_spin} becomes
        \begin{equation} \label{heis_ham}
            \mathcal{H} =  -\frac{J}{2}\sum_{\langle i,j\rangle,\sigma,\sigma'}
            c_{i\sigma}^\dagger c_{j\sigma}c_{j\sigma'}^\dagger c_{i\sigma'} \ + \ \mbox{constant} \ ,
        \end{equation}
        where $c_{i\sigma}^\dagger$ ($c_{i\sigma}$) creates (annihilates) an electron with spin $\sigma$ at site $i$. This transformation embeds the original \emph{spin} Hilbert space into an \emph{electronic} Hilbert space which also contains non-magnetic sites occupied by zero or two electrons. As a result, Eqs \ref{heis_spin} and \ref{heis_ham} are only equivalent on the restricted electronic subspace with half electron filling and no empty sites or double occupancies. This constraint can be enforced exactly by the so-called Gutzwiller projector $P_{G}$. The advantage of this approach is that pairs of fractional $S\!=\!1/2$ quasiparticles (for the original spin Hamiltonian) can be naturally constructed as particle-hole excitations in the electronic space, projected \emph{a-posteriori} by $P_G$ onto spin configurations with exactly one electron per site.

        The quartic electronic operator in Eq.~\ref{heis_ham} is treated by a mean-field decoupling where the averages $\langle c_{i\sigma}^\dagger c_{i\sigma}\rangle$ and $\langle c_{i\sigma}^\dagger c_{j\sigma}\rangle$ are considered. We adopt the following Ans\"atze for their real-space dependencies : $\langle c_{i\sigma}^\dagger c_{i\sigma}\rangle$ corresponds to a staggered N\'eel order parameter (N) and $\langle c_{i\sigma}^\dagger c_{j\sigma}\rangle$ to a staggered flux (SF) threading square plaquettes of the lattice \cite{dmitriev_1996,wen_1996,nayak_2000} (see Supplementary Materials for exact definitions and more details). To each average corresponds a variational parameter whose value is optimized to minimize the energy (Eq.~\ref{heis_spin}) of the \emph{Gutzwiller-projected} state, $\left|\mbox{SF{+}N}\right\rangle=P_G|\psi_{\rm{MF}}\rangle$. The corresponding mean-field electronic ground-state $|\psi_{\rm{MF}}\rangle$ contains empty and doubly occupied sites and reads 
        \begin{equation}
            |\psi_{\rm MF}\rangle = \prod_{\mathbf k\in {\rm MBZ}} 
            \gamma_{\mathbf k\uparrow-}^\dagger \gamma_{\mathbf k\downarrow-}^\dagger|0\rangle \, ,
        \end{equation}
        where $|0\rangle$ is the electron vacuum and where the $\gamma^{(\dagger)}_{\mathbf k\sigma \pm}$ operators are linear combinations of $c_{\mathbf k\sigma}^{(\dagger)}$ operators that diagonalize the mean-field electronic Hamiltonian. The product over the wave-vector $\mathbf k$ is restricted to the magnetic Brillouin zone (MBZ), a result of the antiferromagnetic unit-cell-doubling. Consequently ``$\pm$'' denotes the band index. In the present case of half electron filling, the ``$-$'' band is fully occupied, and there is a finite gap to the empty ``$+$'' band for non-zero N\'eel order-parameter. The overall minimization procedure is carried out numerically using Variational Monte Carlo \cite{gros_1989} and leads to a $\left|{\mbox{SF{+}N}}\right\rangle$ state with variational energy $E_{\rm{SF+N}}=-0.664J$ and staggered moment $0.75S$ per site \cite{gros_1988,lee_1988}. This can be compared to more precise Green's function Monte Carlo studies for Eq.~\ref{heis_spin} that obtained $-0.669J$ and $0.615S$ for the ground-state energy and the staggered moment, respectively \cite{trivedi_1990,buonaura_1998}.

        The optimized $\left|{\mbox{SF{+}N}}\right\rangle$ state, while giving a good estimate for the ground-state energy, does not have the correct  long-distance behaviour for the transverse equal-time correlator $\langle S^+(\mathbf{0})S^-(\mathbf{r})\rangle$, predicted by SWT to decay as a power-law \cite{ogushi_1960}. Instead, as the excitation spectrum of the mean-field electronic ground-state is gapped,  $\langle S^+(\mathbf{0})S^-(\mathbf{r})\rangle$ decays exponentially after projection. We conjecture that the asymptotic behavior of the spin correlator is important for the deconfinement of fractional excitations. To obtain insight into the influence of long-distance spin fluctuations, we consider a distinct variational state, $|\mbox{SF}\rangle$, for which the finite staggered-flux is retained but the N\'eel order is reduced to zero. $|\mbox{SF}\rangle$ is a quantum spin-liquid singlet of variational energy $E_{\rm{SF}}\!=\!-0.638J$ that displays a power-law decay of its transverse spin correlations \cite{paramekanti_2004,ivanov_2006}.

        \begin{figure}[t!]
            \centering
            \includegraphics[width=\linewidth]{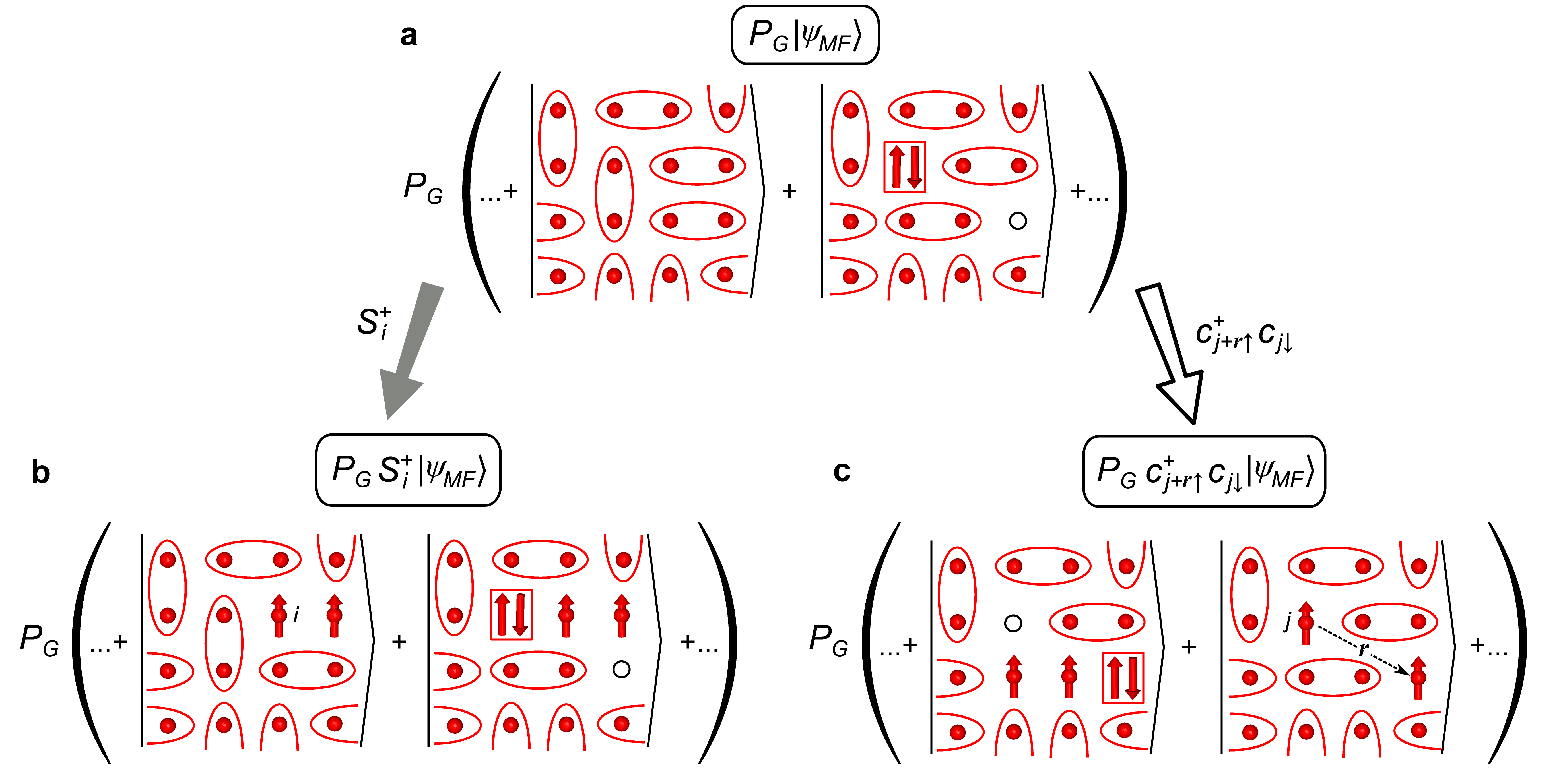}
            \caption{\label{fig3} A schematic representation of local-spin-flip and spatially separated quasiparticle pair excitations in the Gutzwiller-projected approach. ({\bf a}) The mean-field wave function $|\psi_{\rm{MF}}\rangle$ is shown as a resonating-valence-bond liquid (for a better visualization, all singlets are shown as nearest-neighbor and the N\'{e}el order is ignored). Configurations containing doubly occupied sites (right hand side) are discarded by the Gutzwiller projection $P_G$. ({\bf b}) Local spin flips create triplets out of resonating singlets. Configurations from $|\psi_{\rm{MF}}\rangle$ originally containing doubly occupied sites are still projected out (right hand side). ({\bf c}) Nonlocal quasiparticle-pair excitations are constructed as projected particle-hole excitations. At a nonzero separation $\mathbf r$, they contribute by annihilating a doubly occupied site with a hole, leaving two separated spins-$\uparrow$. After projection, the only configurations left are the ones constructed from $|\psi_{\rm{MF}}\rangle$ that contained one empty and one doubly occupied site (right hand side). }
        \end{figure}

        We now turn to the construction of \emph{transverse} ($S\!=\!1$) spin excitations for the above variational states, aiming at comparing their respective dynamic structure factor with the results of Fig.~\ref{fig2}. The variational transverse spin excitations are obtained as superpositions of projected particle-hole excitations with momentum $\mathbf q$:
            \begin{equation}\label{trans_eig}
                |\mathbf q,n,+\rangle=\sum_{\mathbf k\in {\rm MBZ}}\phi_{\mathbf k\mathbf q}^n|\mathbf k,\mathbf q\rangle,\qquad |\mathbf k,\mathbf q\rangle=P_G\gamma_{\mathbf k\uparrow+}^\dagger\gamma_{\mathbf k-\mathbf q\downarrow-}|\psi_{\rm MF}\rangle
            \end{equation}
        where the states $|\mathbf k,\mathbf q\rangle$ are generated by destroying a spin-down quasiparticle in the ``${-}$'' band and creating a spin-up quasiparticle in the ``${+}$'' band. The coefficients $\phi_{\mathbf k\mathbf q}^n$ are obtained by diagonalizing the original Hamiltonian (Eq. \ref{heis_spin}) projected onto the non-orthonormal set of states $|\mathbf k,\mathbf q\rangle$ and correspond to the eigen-energies $E_{\mathbf q n}^+$. Expressing the Fourier-space quasiparticle operators $\gamma_{\mathbf k\sigma \pm}$ using the real-space $c_{i\sigma}$ operators, we note that the variational spin excitations contain both local spin flips $S_i^+ P_G |\psi_{\rm MF}\rangle = P_G c_{i\uparrow}^\dagger c_{i\downarrow} |\psi_{\rm MF}\rangle$ (Fig.~\ref{fig3}{\bf b}) and \emph{spatially separated} particle-hole excitations, $P_G\, c_{j+\mathbf r\uparrow}^\dagger c_{j\downarrow} |\psi_{\rm MF}\rangle$ (Fig.~\ref{fig3}{\bf c}). The dynamic structure factor of the transverse spin excitations is calculated as
        \begin{equation}\label{sqwtrans}
            \mathcal{S}^{+-}(\mathbf q,\omega)=\sum_{n}
            \left|\langle\mathbf{q},n,{+}|S_{\mathbf{q}}^+|\mbox{GS}\rangle\right|^2
            \delta\left(\omega-E_{\mathbf qn}^+ + E_{\rm{GS}}\right)
        \end{equation}
        where $|\mbox{GS}\rangle$ stands either for $|\mbox{SF{+}N}\rangle$ or $|\mbox{SF}\rangle$. We use the identity $\mathcal{S}^\perp\equiv\mathcal{S}^{+-}\!=\!\mathcal{S}^{-+}$ valid for both variational ground-states to compare the transverse dynamic structure factor of the variational states $|\mbox{SF{+}N}\rangle$ and $|\mbox{SF}\rangle$ with the experimental results presented in Fig.~\ref{fig2}. A similar approach also allows to obtain the longitudinal ($S\!=\!0$) dynamic structure factor (see Supplementary Materials).

        \begin{figure}[t!]
            \centering
            \includegraphics[width=\linewidth]{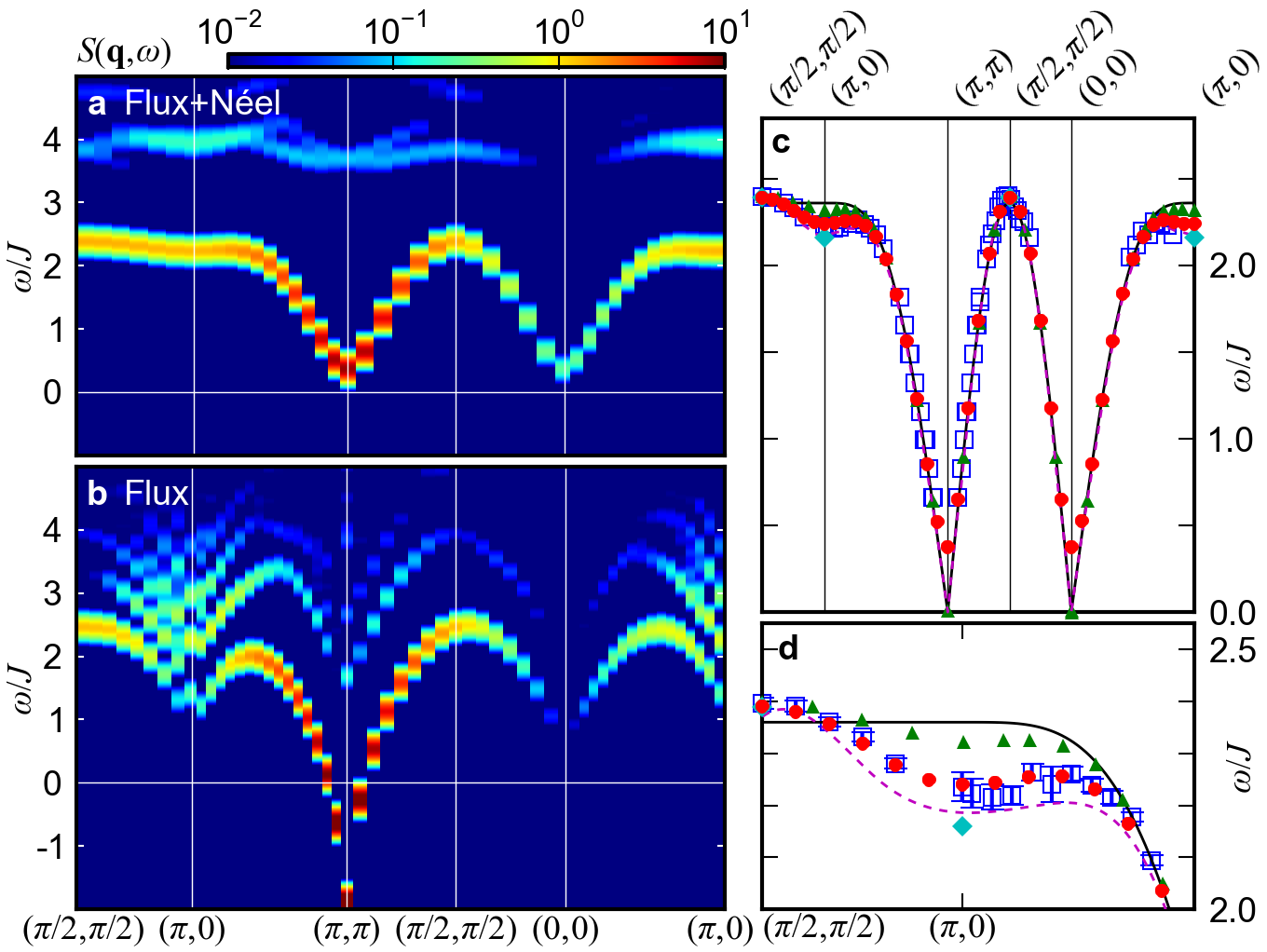}
            \caption{\label{fig4} Calculated transverse dynamic structure factor for the $|\mbox{SF+N}\rangle$ ({\bf a}) and $|\mbox{SF}\rangle$ ({\bf b}) states with lattice sizes of $24\times24$ and $32\times32$ respectively. ({\bf c}) The magnon-like dispersion extracted from {\bf a} (red points) compared to the experimental CFTD data \cite{christensen_2007} (blue squares), spin-wave theory with first-order (solid black line) and third-order \cite{syromyatnikov_2010} (green triangles) $1/S$ corrections, series expansion \cite{zheng_2005} (dashed purple line) and quantum Monte Carlo \cite{sandvik_2001} (cyan diamonds). The experimental data are scaled using $J=6.11$~meV. ({\bf d}) Zoom-in on the magnon-like mode dispersion along the magnetic zone-boundary.}
        \end{figure}

        The transverse dynamic structure factor $\mathcal{S}^{\perp}(\mathbf q,\omega)$ of the $|\mbox{SF{+}N}\rangle$ state is shown in Fig.~\ref{fig4}({\bf a}), as obtained by variational Monte Carlo on a finite lattice of $24\times24$ sites. The dominant feature of the spectrum is a low-energy magnon-like mode, which resembles the  experimental results of Fig.\ \ref{fig1}{\bf a}. In particular, our calculation produces a dispersion along the magnetic zone-boundary in better quantitative agreement with the 7\% dispersion observed in Ref.~\cite{christensen_2007} than any other theoretical method, Figs.\ \ref{fig4}{\bf c} and \ref{fig4}{\bf d}. This confirms that magnons
can be quantitatively interpreted as bound pairs of fractional $S\!=\!1/2$ quasiparticles.

        {However the $|\mbox{SF{+}N}\rangle$ transverse dynamic structure factor exhibits a gap at $(\pi,\pi)$ and no continuum above the magnon branch at $(\pi,0)$. We believe that this is an artifact of replacing the spontaneous symmetry breaking by a N\'eel mean-field order parameter: this ansatz, as mentioned above, distorts the long-distance asymptotics of spin correlations. Indeed, if we reduce the N\'{e}el mean-field parameter of the $|\mbox{SF{+}N}\rangle$ state, then the high-energy excitations at $(\pi,0)$ move down in energy (see Supplementary Materials). At the vanishing N\'{e}el field (i.e., in the $|\mbox{SF}\rangle$ state) they evolve into a succession of modes distributed on an extended energy range above the spin-wave mode (shown in Fig.\ \ref{fig4}{\bf b} for a $32\times32$ lattice). This behavior contrasts the situation at $(\pi/2,\pi/2)$ where the  high-energy transverse excitations completely lose their spectral weight on reducing the N\'eel field and only the spin-wave mode remains in the $|\mbox{SF}\rangle$ state. We therefore suggest that the continuum of excitations observed at $(\pi,0)$ is related to power-law transverse spin correlations and that it corresponds to deconfined fractional spin-1/2 quasiparticles.

        To support this interpretation, we consider in Fig. \ref{fig5}{\bf a} and \ref{fig5}{\bf b} the system-size dependence of $\mathcal{S}^{\perp}(\mathbf q,\omega)$. While the excitations at $(\pi/2,\pi/2)$ form a single sharp mode with energy and intensity nearly independent of the system size, the number of modes at $(\pi,0)$ and their relative  weights are strongly modified by increasing the number of sites. This behavior is consistent with the development of a continuum of fractional quasiparticles at $(\pi,0)$ in the thermodynamic limit. 

        \begin{figure}[t!]
            \centering
            \includegraphics[width=\linewidth]{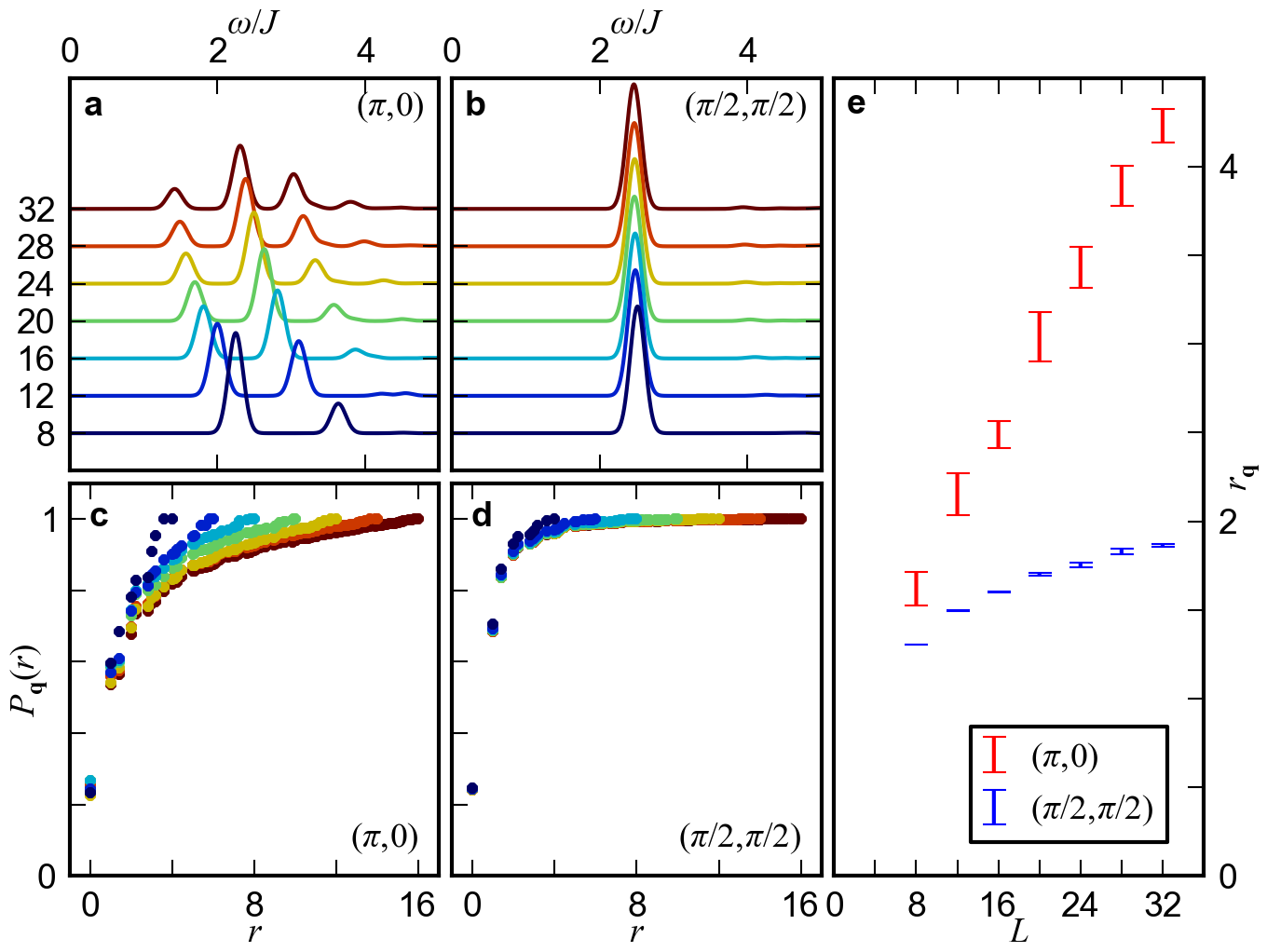}
            \caption{\label{fig5} Finite-size effects in the $|\mbox{SF}\rangle$ state. Transverse dynamic spin structure factor at $(\pi,0)$ ({\bf a})  and $(\pi/2,\pi/2)$ ({\bf b}) for different system sizes ranging from $8\times8$ (dark blue line) to $32\times32$ (dark red line). Disk-integrated fractional-quasiparticle pair separation distribution $P_{\mathbf{q}}(r)$ at $(\pi,0)$ ({\bf c}) and $(\pi/2,\pi/2)$ ({\bf d}) for corresponding system sizes. ({\bf e}) Mean fractional-quasiparticle pair separation $r_{\mathbf{q}}$ at $(\pi,0)$ (red symbols) and $(\pi/2,\pi/2)$ (blue symbols). The error bars result from the variational Monte Carlo sampling.}
        \end{figure}
		
		Having established that our Gutzwiller approach depending on wave-vector produces respectively sharp and continuum-like excitations from the $|\mbox{SF}\rangle$ state, we analyze their \textit{real-space} structure to gain further insight into their nature.} We consider their overlap with projected real-space particle-hole excitations $|\mathbf{q},\mathbf{r}, {+}\rangle=P_G\sum_{\mathbf R}e^{i\mathbf q\cdot\mathbf R}c_{\mathbf R+\mathbf r\uparrow}^\dagger c_{\mathbf R\downarrow}|\psi_{\rm MF}\rangle$, where a Fourier transformation was applied to reflect translation invariance. In this formalism, the most \emph{local} projected particle-hole pair is the spin-flip state $S_{\mathbf q}^+|\mbox{SF}\rangle=|\mathbf q,\mathbf 0,{+}\rangle$ corresponding to a magnon while non-local pairs are characterized by a finite separation $\mathbf r$. Therefore, the degree of deconfinement of a fractional $S\!=\!1/2$ quasiparticles pair can be characterized using the spatial extent of its overlap with projected real-space particle-hole excitations. To evaluate it, we define the weighted average
        \begin{equation}\label{rho}
            \rho_{\mathbf{q}}(\mathbf{r})=\sum_n
            \big| \langle{ \mathbf q,\mathbf r,{+}}| \mathbf{q},n,{+} \rangle \
            \langle {\mathbf q,n,{+}}|S_{\mathbf q}^+|{\rm SF}\rangle \big|^2\, ,
        \end{equation}
        where the aforementioned overlap, represented by the first term, is weighted by the intensity of the transverse spin excitation in the dynamic structure factor. The spatial profile of $\rho_{\mathbf{q}}(\mathbf{r})$ for the magnetic zone-boundary wave-vectors, shown in Figs.\ \ref{fig1}{\bf b} and \ref{fig1}{\bf d}, reveals much more extended fractional $S\!=\!1/2$ quasiparticles pairs at $(\pi,0)$ than at $(\pi/2,\pi/2)$. This is confirmed by the system-size dependence of the radially-integrated normalized distribution $P_{\mathbf q}(r)=\sum_{|\mathbf{r}'|<r}\rho_{\mathbf{q}}(\mathbf{r'})$, plotted in Figs.\ \ref{fig5}{\bf c} and \ref{fig5}{\bf d}. At $(\pi/2,\pi/2)$, $P_{\mathbf q}(r)$ saturates at a distance, $r$, that is nearly independent of the system size, while at $(\pi,0)$ it does so at a distance that \emph{increases} with the number or sites. Similarly, the ``root-mean-square'' fractional quasiparticles pair separation $r_{\mathbf{q}}=\left[\sum_\mathbf{r} |\mathbf{r}|^2 \rho_{\mathbf{q}}(\mathbf{r}) /\sum_\mathbf{r} \rho_{\mathbf{q}}(\mathbf{r})\right]^{1/2}$, presented in Fig.~\ref{fig5}{\bf e}, grows linearly with the system size for $(\pi,0)$ while it has a much weaker size dependence for $(\pi/2,\pi/2)$. 

        Taken together, our real-space results for the $|\mbox{SF}\rangle$ state show that spin excitations at $(\pi/2,\pi/2)$ can indeed be considered as bound pairs of $S=1/2$ quasiparticles with confined spatial extent. In contrast at $(\pi,0)$ the strong system size dependance of the spin excitations spatial extent indicates quasiparticle deconfinement in two spatial dimensions. We do not attempt to extract power laws from the numerical data, since the variational $|\mbox{SF}\rangle$ state mimics the long-distance spin correlations only qualitatively. Similarly, due to the difficulty in recovering the correct asymptotic behavior for the equal-time correlations in the $|\mbox{SF+N}\rangle$ state, it remains an open theoretical challenge to obtain explicit quasiparticle deconfinement out of the magnetically ordered ground state of Eq.~\ref{heis_spin}.

        In conclusion, we have presented experimental and theoretical arguments in favor of the deconfinement of $S\!=\!1$ excitations into
        fractional $S\!=\!1/2$ quasiparticles in the excitation spectrum of the square-lattice quantum Heisenberg antiferromagnet. Within our N\'eel plus Staggered-Flux ansatz, $|\mbox{SF+N}\rangle$, we show that magnons can be interpreted as bound-pairs of fractional quasiparticles. To the best of our knowledge, the results of our Gutzwiller projection approach provide the closest description to date for the high-energy spin-dynamics of the model, as observed experimentally in CFTD and relevant for La$_2$CuO$_4$. Our polarized neutron scattering results unraveled a spin-isotropic excitation continuum at the $(\pi,0)$ wave-vector of the square-lattice Brillouin zone. Similarily, variational Monte-Carlo calculations in the Staggered-Flux quantum spin-liquid ansatz, $|\mbox{SF}\rangle$, reveal an isotropic continuum arising from the spatial deconfinement of fractional quasiparticles pairs. While our theoretical findings \textit{alone} do not unequivocally prove the presence of such deconfinement for the exact ground-state of the model, we argue that \textit{combined} with our experimental results, they provide strong support for the presence of the analogue of deconfined 1D spinons in a two-dimensional non-frustrated long-range ordered antiferromagnet.

        \bibliography{spinons}

        \bibliographystyle{naturemag}

        \section*{Acknowledgments}
        We gratefully acknowledge fruitful discussions with M. Zhitomirsky, S. Sachdev, C. Broholm and L. P. Regnault. Work in EPFL was supported by the Swiss National Science Foundation, the MPBH network, and European Research Council grant CONQUEST. Computational work was supported by the Swiss National Supercomputing Center (CSCS) under project ID s347. Work at Johns Hopkins University was supported in part by the US Department of Energy, office of Basic Energy Sciences, Division of Material Sciences and Engineering under grant DE-FG02-08ER46544. NBC was supported by the Danish Agency for Science, Technology and Innovation under DANSCATT.

        \clearpage


%

        \setcounter{figure}{0}
        \makeatletter
        \makeatletter \renewcommand{\fnum@figure}
        {Supplementary~\figurename~\thefigure}
        \makeatother

        \let\oldthebibliography=\thebibliography
        \let\oldendthebibliography=\endthebibliography
        \renewenvironment{thebibliography}[1]{%
            \oldthebibliography{#1}%
            \setcounter{enumiv}{51}%
        }{\oldendthebibliography}

        \renewcommand{\figurename}{Figure}
        \baselineskip16pt
        \title{\bf Supplementary information for: Fractional excitations in square-lattice quantum antiferromagnet}
        \maketitle
        \tableofcontents
        \clearpage 

        \section{Experimental Materials and Methods}

        \subsection{Properties of the metal-organic compound CFTD}

        A detailed review of the properties of CFTD can be found in Chapter 3 of Ref.~\cite{Christensen2004PhD}.

        \subsubsection{Crystal structure} 
\begin{figure}[t!] 
            \centering
            \includegraphics[width=0.99\linewidth]{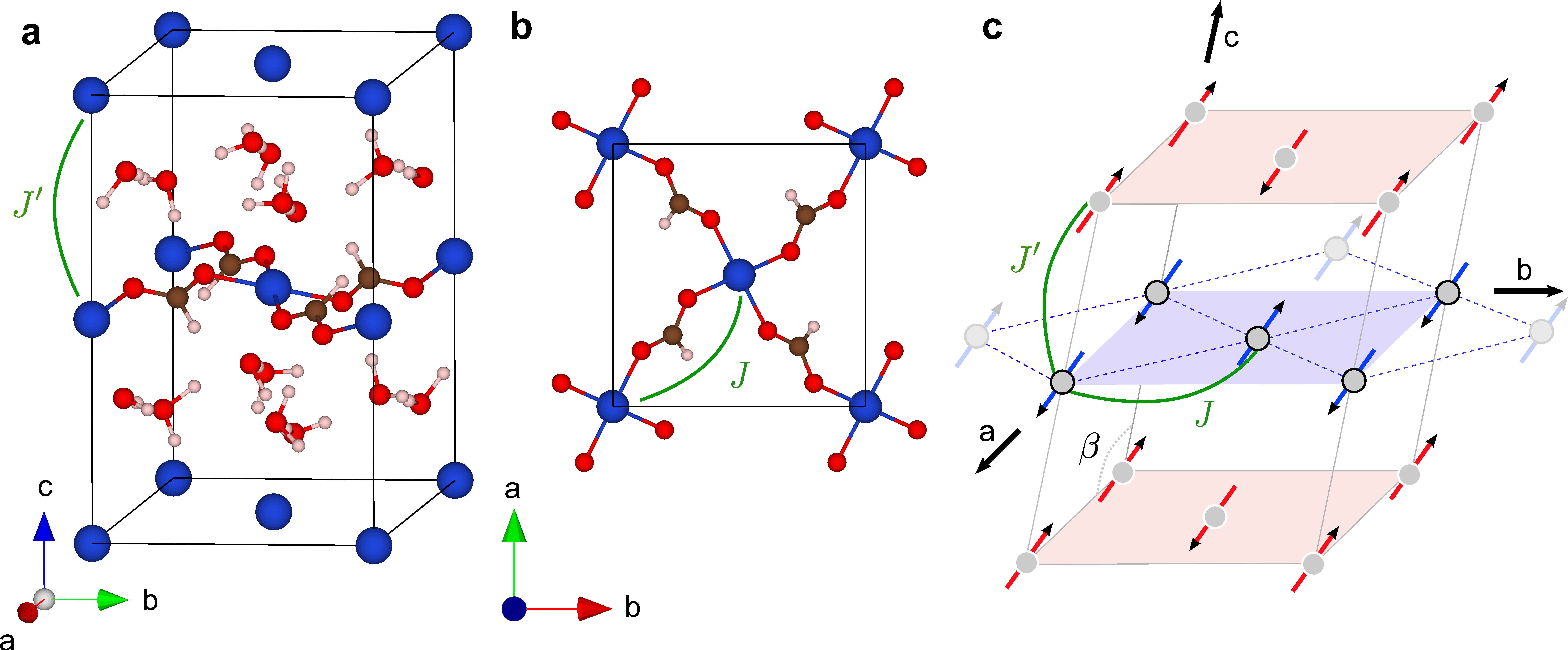}
            \caption{\label{fig12} Crystal and magnetic structure of CFTD. ({\bf a}) Three dimensional crystal structure with Cu, O, C and H/D atoms in blue, red, brown, gray, respectively. Note that the formate groups in the $z=0$ and $z=1$ planes are not represented. ({\bf b}) Two-dimensional copper-formate plane in CFTD and dominant superexchange path $J$ between copper ions. ({\bf c}) Sketch of the magnetic structure of CFTD. Note that the monoclinic angle and the out-of-plane spin direction have been amplified for clarity.}
        \end{figure}

The layered metal-organic compound Cu$($DCOO$)_2\cdot$4D$_2$O (CFTD) investigated in this work is the deuterated analog (hydrogen $^1$H substituted by deuterium $^2$D) of copper formate tetrahydrate (CFTH)~\cite{Kiriyama1954}. It is described in the monoclinic space-group P2$_1/a$ at 300~K and in P2$_1/n$ below $T_s\!\approx\!246$~K where the unit-cell doubles along $c$~\cite{Okada1966}. This change of symmetry results from a first-order antiferroelectric ordering involving the water molecules between the copper-formate planes, see Fig.~S\ref{fig12}{\bf a}. While deuteration influences the temperature of the structural transition $T_s$, CFTH and CFTD are structurally very similar~\cite{Okada1966}. At $T\!=\!120$~K, the lattice parameters of CFTD are $a\!=\!8.113$~\AA, $b\!=\!8.119$~\AA, $c\!=\!12.45$~\AA\ and $\beta = 100^{\circ}$. The two non-equivalent copper sites in the unit-cell are coordinated to four oxygens from formate groups in the $ab$ plane and to two oxygens from crystalline water above and below that plane, forming a staggered pattern of elongated octahedrals. The elongated directions (labeled $\parallel$) are within $\approx\!25^\circ$ of $\boldsymbol{c}$. The two-dimensional lattice of Cu$^{2+}$ ions is face centered with nearest-neighbor coppers arranged on a square lattice (within 0.07\%) at an average distance $d=\frac{1}{2}\sqrt{a^2+b^2}\approx5.74$ \AA, see Fig.~S\ref{fig12}{\bf b}. The monoclinic angle $\beta$ influences the stacking of the $ab$ planes along $\boldsymbol{c}$ but is irrelevant for the two-dimensional physics considered in this work. Assumming the two copper sites indistinguishable, we define the elementary two-dimensional square-lattice by the nearest neighbor vectors $\boldsymbol{x} = (\boldsymbol{a}+\boldsymbol{b})/2d$ and $\boldsymbol{y} = (\boldsymbol{a}-\boldsymbol{b})/2d$ such that $|\boldsymbol{x}|\!=\!|\boldsymbol{y}|=1$.

        \subsubsection{Crystallographic notations} In the reciprocal lattice of CFTD spanned by $\boldsymbol{a}^*$, $\boldsymbol{b}^*$, and $\boldsymbol{c}^*$, we define wave-vectors as ${\bf Q}=h\boldsymbol{a}^*+k\boldsymbol{b}^*+\ell\boldsymbol{c}^*$ with $h$, $k$ and $l$ in reciprocal lattice units (r.l.u). In the $\ell=0$ plane considered in this work, Brillouin zone centers (${\bf Q}=\boldsymbol{\tau}$) are defined by the equivalent reciprocal-lattice vectors  $\boldsymbol{\tau}=(2,0,0)$, $\boldsymbol{\tau}=(1,1,0)$ and $\boldsymbol{\tau}=(0,-2,0)$. In turn, we define the elementary reciprocal square-lattice spanned by $\boldsymbol{x}^*$ and $\boldsymbol{y}^*$ such that the two-dimensional wave-vector ${\bf q}\!=\!q_x\boldsymbol{x}^*+q_y\boldsymbol{y}^*$ with $q_x=\pi(h+k)$ and $q_y=\pi(h-k)$ conform with standard theoretical notations. 

        \subsubsection{Magnetic properties} The magnetic susceptibility of CFTH/CFTD reveals an overall antiferromagnetic behavior with a negative Weiss constant $\Theta_{\rm W}\!=\!-175$~K~\cite{Martin1959}, a 2D short-range order indicated by a broad maximum around $T^*\!\approx\!65$~K \cite{Kobayashi1963}, and a transition to 3D long-range order at $T_{\rm N}\!=\!16.5(5)$~K~\cite{Flippen1963}. The analysis of the $T\!>\!35$~K  susceptibility using high-temperature series-expansion yields a nearest-neighbor exchange $J\!=\!6.2(3)$~meV~\cite{Seehra1969}. Results from local probes such as NMR~\cite{Leeden1967,Dupas1970}, ESR~\cite{Seehra1968} and isothermal magnetization~\cite{Yamagata1980,Yamagata1981} conform with this picture. To fully account for these measurements, it is however necessary to introduce small gyromagnetic and exchange anisotropies. The gyromagnetic-tensor of the Cu$^{2+}$ sites is staggered, with estimated principal components $g_\perp=2.1$ and $g_\parallel=2.4$, and average value of $g_{a\rm v}=2.19$~\cite{Kobayashi1963,Yamagata1980}. In turn, the dominant nearest-neighbor exchange $J$ is weakly anisotropic, with an estimated symmetric off-diagonal component $<\!10^{-3}J$~\cite{Seehra1968} and an estimated antisymmetric component $<\!10^{-2}J$ with Dzyaloshinskii-Moriya vector directed in the $ac$-plane, closer to $\boldsymbol{c}$ than to $\boldsymbol{a}$~\cite{Yamagata1980,Yamagata1981}. The interlayer exchange is estimated to be as small as $J_c\approx5\times10^{-5}J$. Overall, CFTD is one of the best known realization of the spin-1/2 square-lattice Heisenberg antiferromagnet, with an exchange energy-scale perfectly suited for polarized neutron scattering experiment, in contrast to the much higher energy-scale of the parent compounds of the superconducting cuprates.

        \subsubsection{Magnetic order} Unpolarized and polarized neutron diffraction measurements~\cite{Burger1980} establish that CFTD indeed hosts long-range magnetic ordering below $T_{\rm N}$. The static magnetic structure is best described by an anticollinear arrangement of nearest neighbor spins, with spins lying in the $ac$-plane with an on-site moment $\langle{\bf m}\rangle$ of norm $0.48(2)~\mu_{\rm B}$ and direction 8(1)$^\circ$ away from $\boldsymbol{a}$ towards $\boldsymbol{c}$, {\it i.e.} only $3(1)^\circ$ from $\boldsymbol{a}^*$. The four-sublattice structure, see Fig.~S\ref{fig12}{\bf c}, allows a very weak canting along $\boldsymbol{b}$ but a net ferromagnetic contribution $\geq\!0.005(6)~\mu_{\rm B}$ is excluded. This suggests that the magnetic structure is stabilized by a small off-diagonal symmetric exchange with negligible antisymmetric exchange. Only magnetic reflections with $\ell= 2n + 1$ are observed so that there is no magnetic Bragg peak accessible in the $a^*b^*$ reciprocal plane. The lowest-angle magnetic reflection is ${\bf Q}\!=\!(1,0,1)$. As the $\ell$-component of ${\bf Q}$ is irrelevant for the two-dimensional correlations considered in this work, ${\bf Q}\!=\!(1,0,0)$ and equivalent reflections correspond to the center of magnetic Brillouin zone (M-point) with ${\bf q}\!=\!(\pi,\pi)$. Similarly, ${\bf Q}\!=\!(0.5,0.5,0)$ and equivalent positions correspond to the X-point of the magnetic zone-boundary point ${\bf q}=(\pi,0)$, while ${\bf Q}\!=\!(0.5,0,0)$ and equivalent positions coincide with the zone-boundary point ${\bf q}=(\pi/2,\pi/2)$.

        \subsubsection{Magnetic excitation spectrum} Previous inelastic neutron scattering studies on single-crystals of CFTD confirmed it is an excellent realization of the Heisenberg square-lattice antiferromagnet~\cite{Harrison1992,Clarke1999,Ronnow2001,Christensen2004,Christensen2007}. A fit to the dispersion of its magnetic excitations using spin-wave theory, including the renormalization factor $Z_c = 1.18$, yields $J\!=\!6.11(2)$~meV~\cite{Christensen2007}, in good agreement with susceptibility results. Likewise, exchange terms going beyond that of the spin-1/2 square-lattice Heisenberg antiferromagnet are found to be very small. No dispersion is observed along $\boldsymbol{c}^*$~\cite{Harrison1992} what confirms the excellent two-dimensionality of the coumpound. The largest deviation from the ideal model is a small gap at the magnetic-zone center ${\bf Q}=(1,0,0)$ of $\approx\!0.38$~meV~\cite{Clarke1999,Ronnow2001,Christensen2004}, attributed to a small off-diagonal exchange interaction. The absence of further-neighbor interactions is inferred indirectly, by comparing the observed dispersion at the magnetic zone-boundary with theoretical results~\cite{Ronnow2001,Christensen2007}

        \subsection{Crystal Growth}

        Due to the large incoherent neutron cross section of $^1$H, it was necessary to prepare a $^2$D-substituted sample of CFTD for our neutron scattering experiments. This was achieved by dissolving CuO in a solution of D$_2$-formic acid in D$_2$O and evaporating the solution to precipitate CFTD. The product was purified by redissolving, filtering, and recrystallizing twice. Seed crystals were grown by slow evaporation of a supersaturated solution from beakers coated with dimethyldichlorosilane. All operations were carried out under inert atmosphere to avoid exchange with airborne water. 

        The large single crystals used in the experiments were grown by a convection method~\cite{Hope1971}: two glass columns were charged with deuterated starting material and a seed crystal suspended on a glass fiber, respectively, and connected at top and bottom to form a loop. The apparatus was filled with a saturated solution of CFTD in D$_2$O, and a convection current was induced by regulating the temperatures of the starting material column with a water bath. The flow of supersaturated solution generated in the warmer starting material zone over the seed crystal in the cooler zone lead to the growth of a 7.2 g cubic crystal of dimension $20\times20\times20$~mm$^3$ within a period of around 14 days. Three smaller $\approx4$ g crystals were obtained in subsequent growth cycles.

        \subsection{Time-of-flight inelastic neutron scattering experiment}

        Our time-of-flight inelastic neutron scattering results were obtained using the chopper spectrometer MAPS at the ISIS Facility, Rutherford Appleton Laboratory (UK) with an incident neutron energy $E_i\!=\!36.8$ meV. The sample consisted of three co-aligned crystals of total mass $\approx\!12$~g mounted on an aluminum sample holder, aligned with the reciprocal directions $\boldsymbol{a}^*$ and $\boldsymbol{b}^*$ kept perpendicular to the direction of the incident beam, and cooled down to $T\!=\!6$~K in a closed-cycle cryostat. After standard corrections and transformations, the intensity obtained at momentum transfer $\hbar{\bf Q}$ and energy transfer $\hbar\omega$ is proportional to diagonal components of the dynamic structure factor ${\cal S}^{\alpha\alpha}({\mathbf Q},\omega)$ defined as the time Fourier-transform of the thermally averaged spin-spin correlation function,  
        \begin{equation}
            {\cal S}^{\alpha\alpha}({\mathbf Q},\omega)= \frac{1}{2\pi\hbar}\int_{-\infty}^{\infty}dt e^{-i\omega t} \langle S^{\alpha}({\bf Q},0) S^\alpha({\bf Q},t)\rangle_T \ ,
        \end{equation}
        with $S^\alpha({\bf Q},t) =  {1}/{(2\pi)^3}\int S^\alpha({\bf r},t)\exp({i{\bf Q}\cdot{\bf r}}) {d^3{\bf r}}$ the space Fourier transform of the $\alpha$ component of the spin operator ${\bm S}({\bf r},t)$.
        In our time-of-flight experiment ${\alpha\alpha}$ components that are perpendicular to ${\bf Q}$ contribute to the intensity. As an array of detectors is used to cover a large solid angle, each pixel records a different linear combination of ${\alpha\alpha}$-compoments.

        Benefiting from the absence of dispersion of the excitations along $\boldsymbol{c}^*$, the collected data were projected in the $a^*b^*$ reciprocal-plane and integrated along the third momentum transfer direction. Symmetries of the elementary square-lattice were used to fold the data into an irreducible $1/8$ of the two-dimensional Brillouin zone. An energy-dependent background was subtracted from the obtained data. Originating from the imperfect deuteration of our sample due to $^1$H/$^2$D exchange, this coherent and nuclear spin incoherent phonon background is $\bf{Q}$ independent within a good approximation and obtained by integrating on a square $\Delta h$, $\Delta k=\pm0.125$~r.l.u around the nuclear zone center $\boldsymbol{\tau}=(1,1,0)$.

        \subsection{Polarized triple-axis inelastic neutron scattering} 

        \subsubsection{Experimental set-up} Our polarized inelastic neutron scattering results were obtained using the thermal-triple-axis neutron spectrometer IN20 at the Institut Laue-Langevin, Grenoble (France) equipped to perform longitudinal $xyz$ polarization analysis. The sample was mounted on an aluminum sample holder with the reciprocal axes $\boldsymbol{a}^*$ and $\boldsymbol{b}^*$ in the scattering plane and inserted in a standard $^4$He cryostat reaching base temperature $T\!=\!1.5$~K. Polarized neutrons were produced and analysed by horizontally focused Heusler $(111)$ monochromator and analyzer crystals with fixed vertical curvature. The final neutron wave-vector was fixed at $k_f = 2.662$~\AA $^{-1}$. No collimators were installed and the instrument was operated in $W$ configuration. A spin flipper was installed in the scattered beam along with a PG filter to suppress second order contamination. A neutron flux-monitor was installed in $k_i$ to normalize the collected intensity.

        \subsubsection{Polarization analysis} In the above Heusler-Heusler configuration, a set of Helmholtz coils around the sample allowed to provide a guide field of 10-60 Gauss at the sample position, and thus to define the direction of the neutron polarization at the sample. The field and polarization direction is chosen either parallel to the momentum transfer ${\bf Q}$ (${\bm x_0}$), perpendicular to ${\bf Q}$ within the scattering plane (${\bm y_0}$) or perpendicular to the scattering plane, (${\bm z_0}$). Scattering without a change of the neutron spin state (non-spin flip scattering) is denoted accordingly with labels $x_0x_0$, $y_0y_0$, or $z_0z_0$, respectively, while spin-flip scattering is measured by activating a neutron spin-flipper in the outgoing beam after the sample, and labeled $x_0\overline{x_0}$, $y_0\overline{y_0}$, $z_0\overline{z_0}$.

        This way, we measured the following partial cross sections:
        \begin{equation}
            \label{pola}
            \begin{array}{cccccc}
                \frac{{\rm d}^2\sigma}{{\rm d}\Omega {\rm d} E_f}\big|_{x_0\overline{x_0}}&\propto&     & \langle M_{\perp \bm Q}^{y_0{\ast}} M_{\perp \bm Q}^{y_0} \rangle_{T,\omega} &+ \langle M_{\perp \bm Q}^{z_0{\ast}} M_{\perp \bm Q}^{z_0} \rangle_{T,\omega}&+\frac{2}{3} \langle \sigma_{nsi} \rangle_{T,\omega} \\
                \frac{{\rm d}^2\sigma}{{\rm d}\Omega {\rm d} E_f}\big|_{y_0\overline{y_0}}&\propto&     &   &+ \langle M_{\perp \bf Q}^{z_0{\ast}}  M_{\perp \bf Q}^{z_0}  \rangle_{T,\omega}&+\frac{2}{3} \langle \sigma_{nsi} \rangle_{T,\omega} \\
                \frac{{\rm d}^2\sigma}{{\rm d}\Omega {\rm d} E_f}\big|_{z_0\overline{z_0}}&\propto&     &+ \langle M_{\perp \bf Q}^{y_0{\ast}}  M_{\perp \bf Q}^{y_0}  \rangle_{T,\omega}&&+\frac{2}{3} \langle \sigma_{nsi} \rangle_{T,\omega} \\
                           \frac{{\rm d}^2\sigma}{{\rm d}\Omega {\rm d} E_f}\big|_{x_0x_0}&\propto&\langle N^\ast  N  \rangle_{T,\omega} + \langle \nu_{iso}\rangle_{T,\omega} & & & +\frac{1}{3} \langle \sigma_{nsi} \rangle_{T,\omega} \\
                           \frac{{\rm d}^2\sigma}{{\rm d}\Omega {\rm d} E_f}\big|_{y_0y_0}&\propto&\langle N^\ast  N  \rangle_{T,\omega} + \langle \nu_{iso}\rangle_{T,\omega}     & + \langle M_{\perp \bf Q}^{y_0{\ast}}  M_{\perp \bf Q}^{y_0}  \rangle_{T,\omega}& &+\frac{1}{3} \langle \sigma_{nsi} \rangle_{T,\omega} \\
                           \frac{{\rm d}^2\sigma}{{\rm d}\Omega {\rm d} E_f}\big|_{z_0z_0}&\propto&\langle N^\ast  N  \rangle_{T,\omega} + \langle \nu_{iso}\rangle_{T,\omega} & &+ \langle M_{\perp \bf Q}^{z_0{\ast}}  M_{\perp \bf Q}^{z_0}  \rangle_{T,\omega}&+\frac{1}{3} \langle \sigma_{nsi} \rangle_{T,\omega} 
            \end{array}
        \end{equation}
        where $\langle A^{\ast} A \rangle_{T,\omega}$ denotes the time Fourier transformed and thermally averaged pair correlation function, 
        \begin{equation}
            \langle A^{\ast} A \rangle_{T,\omega} = \frac{1}{2\pi}\int_{-\infty}^{\infty}dt e^{-i\omega t} \langle A^{\ast}({\bf Q},0) A({\bf Q},t)\rangle_{T}\ ,
        \end{equation}
        $N(\bf Q)$ is the nuclear scattering amplitude, $\bm M_{\perp \bf Q} (\bf Q)$ the magnetic scattering amplitude perpendicular to the momentum transfer $\bf Q$, $\nu_{iso}$ denotes the nuclear spin-independent elastic and inelastic incoherent scattering due to random isotope distribution, and $\sigma_{nsi}$ the nuclear spin-dependent elastic and inelastic incoherent scattering related to random nuclear spin orientation. The proportionality constant is the same for all six equations. From the above six equations, the four quantities $\langle M_{\perp \bf Q}^{y_0{\ast}} M_{\perp \bf Q}^{y_0} \rangle_{T,\omega}$, $\langle M_{\perp \bf Q}^{z_0{\ast}} M_{\perp \bf Q}^{z_0} \rangle_{T,\omega}$, $\langle \sigma_{nsi} \rangle_{T,\omega}$, and $\langle N^\ast  N  \rangle_{T,\omega} + \langle \nu_{iso}\rangle_{T,\omega}$ can be determined separately. 

        \begin{figure}[t!]
            \centering
            \includegraphics[width=0.85\linewidth]{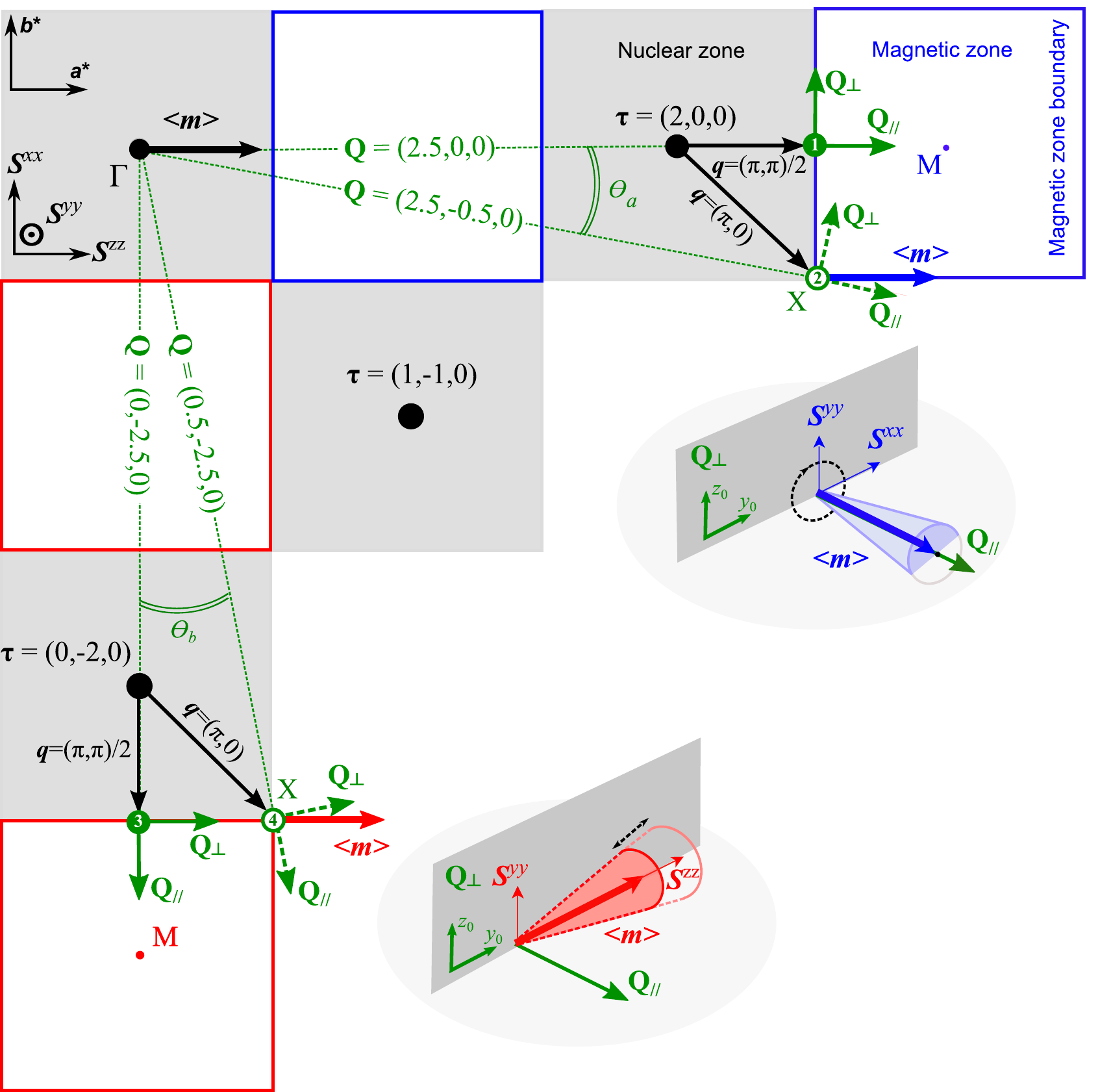}
            \caption{\label{fig11} Representation of the $a^*b^*$ reciprocal plane of CFTD. The gray areas represent the nuclear zones (center $\Gamma$) while the white areas are the magnetic zones (center M). The magnetic zone-boundaries are represented as thick solid (red and blue) squares. The wave-vectors ${\bf Q}$ investigated in our triple-axis experiment are represented by green circles along with their corresponding 2D wave-vector ${\bf q}$ from the closest nuclear zone center $\boldsymbol{\tau}$. The direction of the ordered moment ($3(1)^\circ$ from $\boldsymbol{a}^*$ towards $\boldsymbol{c}$) is represented as a bold arrow. The ``transverse'' and ``longitudinal'' configurations with respect to the ordered moment are schematically drawn in blue and red, respectively.}
        \end{figure}

        \subsubsection{Reconstruction of the dynamic structure factor}

The components of the magnetic vector $M^\alpha({\bf Q},t)$ entering the polarization analysis formulas of Eq.~S\ref{pola} are related to the space Fourier-transform of spin operators, $M^\alpha({\bf Q},t)= \frac{1}{2}g f({\bf Q}) S^\alpha({\bf Q},t)$, where $g$ is the the gyromagnetic factor and $f({\bf Q})$ the magnetic form-factor of the considered $\rm Cu^{2+}$ ions. Polarization analysis thus allows to extract separately two components of the dynamic structure factor at each ${\bf Q}$. 

        Note that the $\alpha=x, y, z$ components of ${\cal S}^{\alpha\alpha}({\mathbf Q},\omega)$ are expressed in the Cartesian coordinate system attached to the ordered-moment direction $\boldsymbol{z}\parallel\langle{\bf m}\rangle$ while the $\beta=y_0, z_0$ components of $\bm M_{\perp \bf Q} (\bf Q)$ are expressed in the Cartesian coordinate system attached to the momentum transfer $\boldsymbol{x}_0\parallel{\bf Q}$. This allows to reconstruct the full dynamic structure factor by combining measurements from two equivalent Brillouin Zones, the orientations of which with respect to $\langle{\bf m}\rangle$ differs, see Fig.~{S}\ref{fig11}. In the $\boldsymbol{\tau}=(2,0,0)$ zone, ${\bf Q}$ is mostly parallel to $\langle{\bf m}\rangle$ what allows to obtain the transverse components ${\cal S}^{xx}({\mathbf Q},\omega)$ and ${\cal S}^{yy}({\mathbf Q},\omega)$ by measuring ${\bf Q}_1=(2.5,0,0)$ and ${\bf Q}_2=(2.5,-0.5,0)$ corresponding to the 2D wave-vectors ${\bf q}=(\pi/2,\pi/2)$ and ${\bf q}=(\pi,0)$, respectively. Likewise, in the $\boldsymbol{\tau}=(0,-2,0)$ zone, ${\bf Q}$ is mostly perpendicular to $\langle{\bf m}\rangle$, what allows to obtain the longitudinal and transverse components ${\cal S}^{zz}({\mathbf Q},\omega)$ and ${\cal S}^{yy}({\mathbf Q},\omega)$ for ${\bf Q}_3=(0,-2.5,0)$ and ${\bf Q}_4=(0.5,-2.5,0)$ corresponding to the 2D wave-vectors ${\bf q}=(\pi/2,\pi/2)$ and ${\bf q}=(\pi,0)$, respectively. This is conveniently understood schematically, see Fig.~{S}\ref{fig11}.

        \subsubsection{Data correction} Our experiment thus consisted in collecting the six above partial cross sections of Eq.~S\ref{pola} as a function of energy transfer for the four momentum transfers of Fig.~S\ref{fig11}. Various corrections had to be applied to relate the measured intensities to the components of the dynamic structure factor. First, the intrinsic $k_i^{-1}$ sensitivity of the flux monitor, corrected for higher-order neutron contamination, balanced the $k_i$ dependence of the neutron scattering cross-section. In addition, a geometrical correction was applied to account for the variation in the vertical focal length at the sample position as a function of incident neutron  energy. Then, due to the imperfect beam polarization, the data were corrected for the mutual influence of non-spin-flip and spin-flip scattering into another in each channel,
        \begin{equation}
            \left( \begin{array}{c}
                \frac{{\rm d}^2\sigma}{{\rm d}\Omega {\rm d}E_f}\big|_{\beta\bar{\beta}} \vspace{0.2cm} \\
                \frac{{\rm d}^2\sigma}{{\rm d}\Omega {\rm d}E_f}\big|_{\beta\beta}  
            \end{array} 
            \right) = \frac{1}{2}
            \left( \begin{array}{cc}
                1+P_{\beta} & \hspace{0.2cm} 1-P_{\beta} \vspace{0.2cm} \\
                1-P_{\beta} & \hspace{0.2cm} 1+P_{\beta}
            \end{array} 
            \right) 
            \left( \begin{array}{c}
                \frac{{\rm d}^2\sigma}{{\rm d}\Omega {\rm d}E_f}\big|^{\prime}_{\beta\bar{\beta}} \vspace{0.2cm} \\
                \frac{{\rm d}^2\sigma}{{\rm d}\Omega {\rm d}E_f}\big|^{\prime}_{\beta\beta} 
            \end{array} 
            \right)
        \end{equation}
        where $P_{\beta}$ is the beam polarization in the direction $\beta$, ${{\rm d}^2\sigma}/{{\rm d}\Omega {\rm d}E_f}$ is a measured cross-section and ${{{\rm d}^2\sigma}/{{\rm d}\Omega {\rm d}E_f}}\large|^{\prime}$ a corrected one. Following, careful tests performed during the experiment on the polarization of the direct-beam, we used an isotropic imperfect polarization $P_{x_0}\!=\!P_{y_0}\!=\!P_{z_0}\!=\!0.873(5)$. Given the small anisotropy of the gyromagnetic factor and noting that all wave-vectors considered  here have similar length within $2\%$, we obtain that the magnetic form factor $g f({\bf Q}_i)/2$ is constant. We also neglected the difference in extinction between our four ${\bf Q}_i$ given the nearly cubic shape of the sample and also neglected the $3^\circ$ out-of-plane orientation of the ordered moment.

        \subsubsection{Results} Finally, we obtain $\langle M_{\perp \bf Q}^{y_0{\ast}}M_{\perp \bf Q}^{y_0} \rangle_{_{T}} ({\bf Q},\omega)$ and $\langle M_{\perp \bf Q}^{z_0{\ast}} M_{\perp \bf Q}^{z_0} \rangle_{_{T}} ({\bf Q},\omega)$ after correcting the measured cross-sections using Eq.~S\ref{pola}, and reconstruct the full dynamic structure factor using  the equations below. For the ${\bf q} = (\pi/2,\pi/2)$ wave-vector, we reconstruct the dynamic structure factor using:
        \begin{eqnarray}
            \mathcal{S}^{zz}({\bf q},\omega) &=& 
            \langle M_{\perp \bf Q}^{y_0{\ast}} M_{\perp \bf Q}^{y_0} \rangle_{_{T}}({\bf Q}_3,\omega) \nonumber  \ , \\[6pt]
            \mathcal{S}^{xx}({\bf q},\omega) &=&    
            \langle M_{\perp \bf Q}^{y_0{\ast}} M_{\perp \bf Q}^{y_0} \rangle_{_{T}}({\bf Q}_1,\omega)\ ,    \\
            \mathcal{S}^{yy}({\bf q},\omega) &=&    \frac{1}{2} \big[
            \langle M_{\perp \bf Q}^{z_0{\ast}} M_{\perp \bf Q}^{z_0} \rangle_{_{T}}({\bf Q}_1,\omega) +    
        \langle M_{\perp \bf Q}^{z_0{\ast}} M_{\perp \bf Q}^{z_0} \rangle_{_{T}}({\bf Q}_3,\omega)\big] \ .     \nonumber 
    \end{eqnarray}
    For the ${\bf q} = (\pi,0)$ wave-vector, we reconstruct the dynamic structure factor using
    \begin{eqnarray}
        \left( \begin{array}{l}
            \mathcal{S}^{zz}({\bf q},\omega) \vspace{0.2cm} \\
            \mathcal{S}^{xx}({\bf q},\omega) 
        \end{array} \right) &= &
            \left( \begin{array}{cc}
                \sin^2\theta_a & \cos^2\theta_a \vspace{0.2cm} \\
                \cos^2\theta_b & \sin^2\theta_b
            \end{array} \right)^{-1} 
            \left( \begin{array}{c}
                \langle M_{\perp \bf Q}^{y_0{\ast}} M_{\perp \bf Q}^{y_0} \rangle_{_{T}}({\bf Q}_2,\omega)\vspace{0.2cm} \\
                \langle M_{\perp \bf Q}^{y_0{\ast}} M_{\perp \bf Q}^{y_0} \rangle_{_{T}}({\bf Q}_4,\omega)
            \end{array} \right) \ , \\
            \mathcal{S}^{yy}({\bf q},\omega) &=&    \frac{1}{2} \big[
            \langle M_{\perp \bf Q}^{z_0{\ast}} M_{\perp \bf Q}^{z_0} \rangle_{_{T}}({\bf Q}_2,\omega) +    
        \langle M_{\perp \bf Q}^{z_0{\ast}} M_{\perp \bf Q}^{z_0} \rangle_{_{T}}({\bf Q}_4,\omega)\big] \ ,\nonumber 
    \end{eqnarray}
    where $\theta_a \approx \theta_b \approx 10^\circ$ accounts for the angle between ${\bf Q}$ and the directions of the reciprocal space vectors $\boldsymbol{a}^*$ and $\boldsymbol{b}^*$, respectively. In the end, the transverse, longitudinal and total contributions to the dynamic structure factor are obtained as,
    \begin{eqnarray}
        \mathcal{S}^{\pm}({\bf q},\omega)&=&\frac{1}{2}[\mathcal{S}^{+-}({\bf q},\omega)+\mathcal{S}^{-+}({\bf q},\omega)] = \mathcal{S}^{xx}({\bf q},\omega)+\mathcal{S}^{yy}({\bf q},\omega), \nonumber \\ 
          \mathcal{S}^{L}({\bf q},\omega)&=&\mathcal{S}^{zz}({\bf q},\omega), \\[6pt] \nonumber
              \mathcal{S}({\bf q},\omega)&=&\mathcal{S}^{xx}({\bf q},\omega)+\mathcal{S}^{yy}({\bf q},\omega)+\mathcal{S}^{zz}({\bf q},\omega) \ .
    \end{eqnarray}

    \section{Theoretical Materials and Methods}

    \subsection{SF+N mean-field Hamiltonian}

    The SF+N Hamiltonian is a mean-field ansatz for the Heisenberg model expressed in terms of fermionic operators (Eq.\ \ref{heis_ham}). The mean-field decoupling is:
    \begin{equation}
        \mathcal{H}_{\rm{MF}}=\sum_{\langle i,j\rangle,\sigma}\left(\chi_{ij}c_{i\sigma}^\dagger c_{j\sigma}+\rm{h.c}\right) + \sum_{i\sigma}(-1)^ih_\sigma c_{i\sigma}^\dagger c_{i\sigma}
    \end{equation}
    with
    \begin{equation}
        h_\sigma=-\sigma H_{\rm N}\qquad \sigma\in\left\{-\frac{1}{2},\frac{1}{2}\right\}
    \end{equation}
    and
    \begin{equation}
        \chi_{ij}=e^{i\theta_{ij}}\qquad \theta_{ij}=(-1)^{i_x+j_y}\theta_0
    \end{equation}
    for nearest-neighbor pairs $\langle i,j\rangle$, $i_{x(y)}$ being the $x$ ($y$) coordinate of site $i$ on the square lattice. For constructing the ground-state wave function, we ignore the overall scale of the Hamiltonian $\mathcal{H}_{\rm{MF}}$ and put $|\chi_{ij}|=1$. The two variational parameters are $\theta_0$ and $H_{\rm{N}}$ (the N\'eel field). The complex hopping amplitude $\chi_{ij}$ causes the square lattice to be threaded with staggered fluxes going through the lattice squares. Electrons circulating around a square acquire a phase of $\pm4\theta_0$.

    The mean-field Hamiltonian $\mathcal{H}_{\rm{MF}}$ is diagonalized by the quasiparticle operators
    \begin{eqnarray}
        \gamma_{\mathbf k\sigma -}&=&
        u_{\mathbf k\sigma -}\frac{1}{\sqrt{2}}\left(c_{\mathbf k\,\sigma}+c_{\mathbf k+\mathbf{\Pi}\,\sigma}\right)
        +v_{\mathbf k\sigma -}^*\frac{1}{\sqrt{2}}\left(c_{\mathbf k\,\sigma}-c_{\mathbf k+\mathbf{\Pi}\,\sigma}\right)\, ,
        \nonumber \\
        \gamma_{\mathbf k\sigma +}&=&
        u_{\mathbf k\sigma +}^*\frac{1}{\sqrt{2}}\left(c_{\mathbf k\,\sigma}+c_{\mathbf k+\mathbf{\Pi}\,\sigma}\right)
        +v_{\mathbf k\sigma +}\frac{1}{\sqrt{2}}\left(c_{\mathbf k\,\sigma}-c_{\mathbf k+\mathbf{\Pi}\,\sigma}\right)\, ,
    \end{eqnarray}
    where the wave vector $\mathbf k$ is restricted to the magnetic Brillouin zone, $\sigma$ is the spin index, and $\mathbf{\Pi}=(\pi,\pi)$. The coefficients $u_{\mathbf k\sigma\pm}$ and $v_{\mathbf k\sigma\pm}$ are calculated as 
    \begin{eqnarray}
        u_{\mathbf k\sigma-}=v_{\mathbf k\sigma+}&=&\sqrt{\frac{1}{2}\left(1+\frac{\sigma H_{\rm{N}}}{\omega_{\mathbf k}}\right)}\, ,
        \nonumber \\
        v_{\mathbf k\sigma-}=-u_{\mathbf k\sigma+}^*&=&\frac{\Delta_{\mathbf k}}{|\Delta_{\mathbf k}|}\sqrt{\frac{1}{2}\left(1-\frac{\sigma H_{\rm{N}}}{\omega_{\mathbf k}}\right)}\, .
    \end{eqnarray}
    Here $\pm$ subscripts correspond to the upper and lower branch of quasiparticles with energies $\pm\omega_{\mathbf k}$, respectively. Note that these energies do not carry any physical meaning, but only determine the empty and occupied quasiparticle states. They are given by
    \begin{equation}
        \omega_{\mathbf k}=\sqrt{|\Delta_{\mathbf k}|^2+H_{\rm{N}}^2}\, ,
        \qquad 
        \Delta_{\mathbf k}=\frac{1}{2}\left(e^{i\theta_0}\cos k_x+e^{-i\theta_0}\cos k_y\right)\, .
    \end{equation}
    The variational parameters $\theta_0$ and $H_{\rm N}$ are optimized using the standard variational Monte Carlo technique. The optimal wavefunction $|\rm SF+N\rangle$ is obtained with
    \begin{eqnarray}
        \theta_0&=&0.1\pi\\
       H_{\rm N}&=&0.11.
    \end{eqnarray}
    On the other hand, the spin liquid state $|\rm SF\rangle$ is obtained by letting the N\'eel field go to zero
    \begin{eqnarray}
        \theta_0&=&0.1\pi\\
       H_{\rm N}&=&0.
    \end{eqnarray}

    \subsection{Numerical projection of the Heisenberg Hamiltonian}

    In our variational approach, we project the physical Heisenberg spin Hamiltonian $\mathcal{H}$ (given by Eq.\ \ref{heis_spin}) onto the subspace spanned by the Gutzwiller-projected mean-field particle-hole states $|\mathbf k,\mathbf q\rangle$ (Eq. \ref{trans_eig}). For simplicity, in this section we only discuss the case of the transverse excitations; the longitudinal case is treated in a similar way.

    Diagonalizing the projected Hamiltonian amounts to solving a generalized eigenvalue problem
    \begin{equation}
        \sum_{\mathbf k'}\mathcal{H}_{\mathbf k\mathbf k'}^{\mathbf q}\phi_{\mathbf k'\mathbf q}^n = 
        E_{\mathbf q n}^+ \sum_{\mathbf k'}\mathcal{O}_{\mathbf k\mathbf k'}^{\mathbf q} \phi_{\mathbf k'\mathbf q}^n
        \label{eig_prob}
    \end{equation}
    in order to determine the coefficients $\phi_{\mathbf k'\mathbf q}^n$ of the expansion of the excitations in the basis $|\mathbf k,\mathbf q\rangle$ (Eq.\ \ref{trans_eig}) and the excitation energies $E_{\mathbf q n}^+$. The matrices $\mathcal{H}_{\mathbf k\mathbf k'}^{\mathbf q}$ and $\mathcal{O}_{\mathbf k\mathbf k'}^{\mathbf q}$ are defined as
    \begin{equation}
        \mathcal{H}_{\mathbf k\mathbf k'}^{\mathbf q} = \langle \mathbf k,\mathbf q | \mathcal{H} | \mathbf k',\mathbf q \rangle\, ,
        \qquad
        \mathcal{O}_{\mathbf k\mathbf k'}^{\mathbf q} = \langle \mathbf k,\mathbf q | \mathbf k',\mathbf q \rangle\, .
    \end{equation}
    These matrices are sampled using the Monte Carlo reweighing technique developed by Li and Yang\cite{li_2010}:
    \begin{eqnarray}
        \mathcal{H}_{\mathbf k\mathbf k'}^{\mathbf q} &=&  
        \mathrm{Tr}(\mathcal{O}^{\mathbf q})
        \sum_{\alpha}\underbrace{\frac{|\langle \mathbf k,\mathbf q|\alpha\rangle|^2}{
        \mathrm{Tr}(\mathcal{O}^{\mathbf q})}}_{\rho_{\mathbf q}(\alpha)}
        \sum_\beta\frac{\langle\mathbf k,\mathbf q|\alpha\rangle\langle\alpha|\mathcal{H}|
        \beta\rangle\langle\beta|\mathbf k',\mathbf q\rangle}{\sum_{\mathbf k^{''}}|\langle \mathbf k^{''},\mathbf q|\alpha\rangle|^2}\, , 
        \nonumber\\
        \mathcal{O}_{\mathbf k\mathbf k'}^{\mathbf q} &=&
        \mathrm{Tr}(\mathcal{O}^{\mathbf q})
        \sum_{\alpha}\rho_{\mathbf q}(\alpha)\frac{\langle\mathbf k,\mathbf q|\alpha\rangle\langle\alpha|\mathbf k',\mathbf q\rangle}{\sum_{\mathbf k^{''}}|\langle \mathbf k^{''},\mathbf q|\alpha\rangle|^2}\, .
        \label{H_O_projected}
    \end{eqnarray}
    Here $|\alpha\rangle$ and $|\beta\rangle$ are real space spin configurations and $\rho_{\mathbf q}(\alpha)$ are the Monte Carlo sampling probabilities. The amplitudes $\langle\mathbf k,\mathbf q|\alpha\rangle$ are calculated as Slater determinants \cite{gros_1989}. The amplitudes $\langle\beta|\mathbf k',\mathbf q\rangle$ are efficiently calculated by changing a row and a column in the Slater matrix using a rank-2 determinant update. Note that this approach leaves uncalculated the overall normalization factor $\mathrm{Tr}(\mathcal{O}^{\mathbf q})=\sum_{\mathbf k}\mathcal{O}^{\mathbf q}_{\mathbf k\mathbf k}$ in Eq.\ S\ref{H_O_projected}, which can be ignored for the generalized eigenvalue problem (Eq.\ S\ref{eig_prob}).


    \subsection{Transverse dynamic structure factor}

    Since the local spin-flip operator $S_i^+$ commutes with the Gutzwiller projection, its action on the variational ground state $|{\rm SF({+}N)}\rangle$ can be decomposed in the projected particle-hole states $|\mathbf k,\mathbf q\rangle$:
    \begin{equation}
        S_{\mathbf q}^+|{\rm SF({+}N)}\rangle=\sum_{\mathbf k}\tilde\phi_{\mathbf k\mathbf q}|\mathbf k,\mathbf q\rangle\, ,
    \end{equation}
    with the expansion coefficients
    \begin{equation}
        \tilde\phi_{\mathbf k\mathbf q}=u_{\mathbf k\uparrow+}^*u_{\mathbf k-\mathbf q\downarrow-}
        +v_{\mathbf k\uparrow+}^*v_{\mathbf k-\mathbf q\downarrow-}\, .
    \end{equation}
    More generally, a projected transverse particle-hole pair state can be written as
    \begin{equation}
        |\mathbf{q},\mathbf{r},{+}\rangle=P_G\sum_{\mathbf{R}}
        e^{i\mathbf{q}\cdot\mathbf{R}} c_{\mathbf{R}+\mathbf{r}\uparrow}^\dagger c_{\mathbf{R}\downarrow}|\psi_{\rm{MF}}\rangle
        =\sum_{\mathbf{k}}\tilde\phi_{\mathbf{k}\mathbf{q}}(\mathbf{r})|\mathbf{k},\mathbf{q}\rangle
    \end{equation}
    with
    \begin{equation}
        \tilde\phi_{\mathbf{k}\mathbf{q}}(\mathbf{r})=e^{-i\mathbf{k}\cdot\mathbf{r}}\left[ \epsilon_{\mathbf{r}}
            \left(u_{\mathbf k\uparrow+}^*u_{\mathbf k-\mathbf q\downarrow-}+v_{\mathbf k\uparrow+}^*
            v_{\mathbf k-\mathbf q\downarrow-}\right) +\bar{\epsilon}_{\mathbf{r}}\left(v_{\mathbf k\uparrow+}^*
        u_{\mathbf k-\mathbf q\downarrow-}+u_{\mathbf k\uparrow+}^*v_{\mathbf k-\mathbf q\downarrow-}\right)\right]\, ,
    \end{equation}
    $\epsilon_{\mathbf{r}}=\frac{1}{2}\left(1+e^{i\mathbf{\Pi}\cdot\mathbf{r}}\right)$, and $\bar{\epsilon}_{\mathbf{r}}=\frac{1}{2}\left(1-e^{i\mathbf{\Pi}\cdot\mathbf{r}}\right)$. The zero-temperature transverse dynamic structure factor of a system in the groundstate $|\psi_{\rm GS}\rangle$ (with the ground-state energy $E_{\rm GS}$) is
    \begin{equation}
        \mathcal{S}^{\pm}(\mathbf q,\omega)=\sum_{\lambda_{\mathbf q}}
        \langle \psi_{\rm GS}|S_{\mathbf q}^-|\lambda_{\mathbf q}\rangle\langle\lambda_{\mathbf q}|S_{\mathbf q}^+|\psi_{\rm{GS}}\rangle\delta(\omega-E_{\lambda_{\mathbf q}}+E_{\rm GS})\, ,
    \end{equation}
    where $\{|\lambda_{\mathbf q}\rangle\}$ is the set of all excited states with energies $E_{\lambda_{\mathbf q}}$. In our variational approximation, we restrict the sum to the projected particle-hole eigen-states as obtained by solving the generalized eigenvalue problem (Eq.\ S\ref{eig_prob}) and obtain:
    \begin{equation}
        \mathcal{S}^{\pm}(\mathbf q,\omega)=\sum_n\left|\sum_{\mathbf k,\mathbf k'}\phi_{\mathbf k\mathbf q}^{n*}
        \mathcal{O}_{\mathbf k\mathbf k'}^{\mathbf q}\tilde\phi_{\mathbf k'\mathbf q}\right|^2
        \delta(\omega-E_{\mathbf q n}^++E_{\rm{SF(+N)}})\, .
    \end{equation}
    Note that in order to calculate this expression, we need to know the normalization of the overlap matrix $\mathcal{O}_{\mathbf k\mathbf k'}^{\mathbf q}$. This normalization was disregarded in the generalized-eigenvalue calculation (Eq.\ S\ref{H_O_projected}), but can be calculated independently from the sum rule
    \begin{equation}
        \langle S_{\mathbf q}^-S_{\mathbf q}^+\rangle = \int d\omega \mathcal{S}^\pm(\mathbf q,\omega)
        = \sum_n\left|\sum_{\mathbf k,\mathbf k'}\phi_{\mathbf k\mathbf q}^{n*}
        \mathcal{O}_{\mathbf k\mathbf k'}^{\mathbf q}\tilde\phi_{\mathbf k'\mathbf q}\right|^2
        =\sum_{\mathbf k,\mathbf k'}\tilde\phi_{\mathbf k'\mathbf q}^{*}\mathcal{O}_{\mathbf k\mathbf k'}^{\mathbf q}
        \tilde\phi_{\mathbf k'\mathbf q}\, .
    \end{equation}
    This is the equal-time transverse spin correlation function, which can be calculated directly as an expectation value in the variational ground state by using a Monte Carlo sampling. This method allows us to calculate the correctly normalized transverse dynamic structure factor.


    \begin{figure}[t!]
        \centering
        \includegraphics[width=0.8\linewidth]{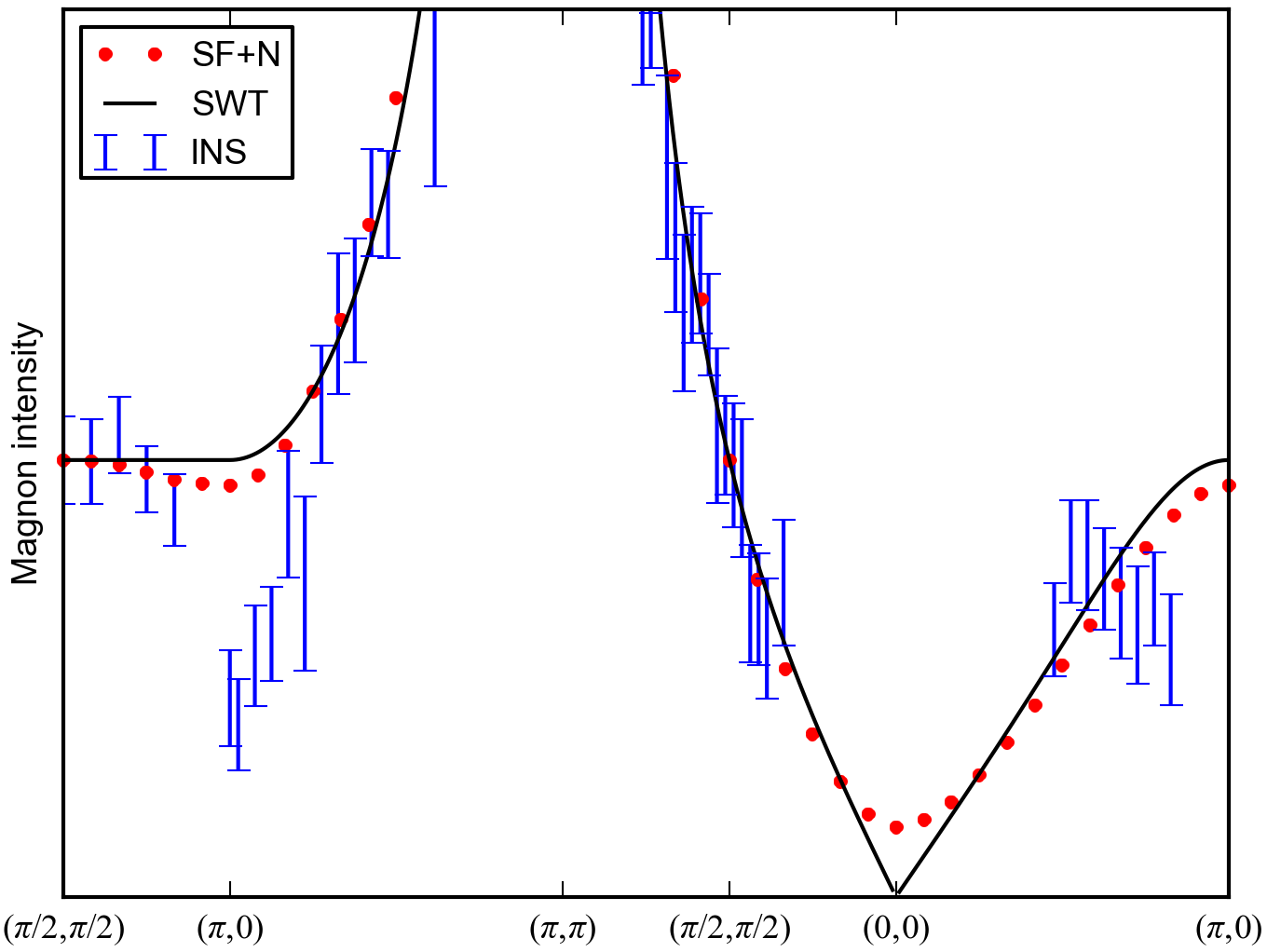}
        \caption{\label{fig6} Magnon intensity along the high-symmetry directions of the Brillouin zone. Solid black line: linear spin-wave theory. Red dots: Lowest energy exciations of the $|\mbox{SF{+}N}\rangle$ variational state. Blue errorbars: experimental results from ref. \cite{christensen_2007}. The $|\mbox{SF{+}N}\rangle$ and linear spin-wave theory intensities are scaled such that they coincide with the experimental data at $\mathbf{q}=(\pi/2,\pi/2)$.}
    \end{figure}

    \subsection{Transverse magnon mode intensity}

    We compare here our variational calculation on the $|\mbox{SF{+}N}\rangle$ state with SWT. Remarkably, it reproduces well not only the energies of the magnon excitations, but also the intensities \begin{equation} I_{\mathbf{q},n}=\left|\langle\mathbf{q},n|S_{\mathbf{q}}^+|\psi_{\rm{GS}}\rangle\right|^2\, , \end{equation} except for the low-energy Goldstone magnons and for the $(\pi,0)$ point (where some part of the calculated spectral weight is transferred to higher energy states), see Fig.\ S\ref{fig6}. However both theories fail to capture the dramatic loss of spectral weight observed at $(\pi,0)$ in experiments, as explained in the main text.


    \begin{figure}[t!]
        \centering
        \includegraphics[width=0.8\linewidth]{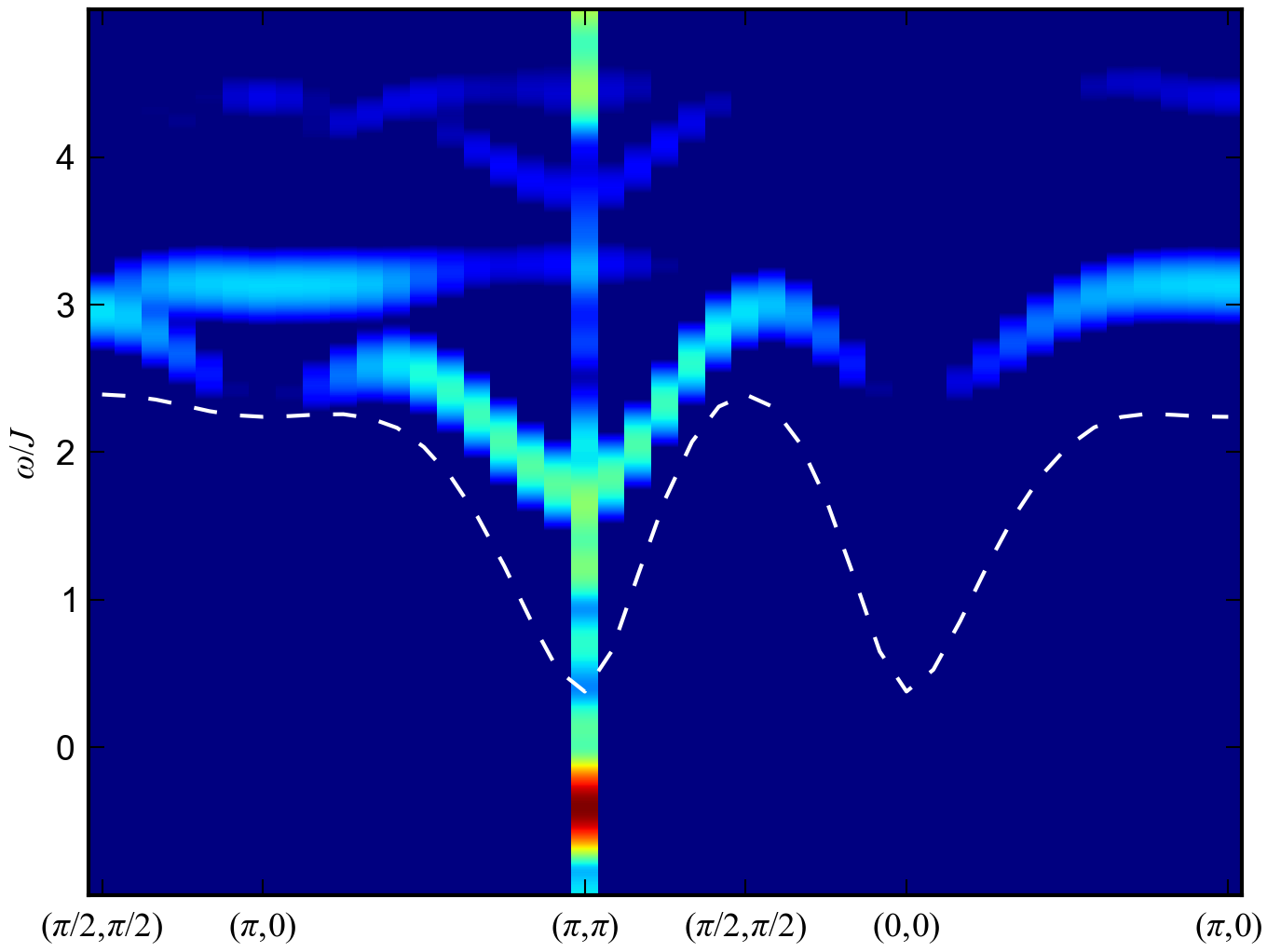}
        \caption{\label{fig7} The longitudinal dynamic structure factor of the $|\mbox{SF{+}N}\rangle$ state for the system size $24\times24$. The dashed white line shows the energies of the magnon mode found in the transverse dynamic structure factor.}
    \end{figure}

    \subsection{Longitudinal dynamic structure factor}

    To obtain the longitudinal dynamic structure factor, we construct a set of $\Delta S_z=0$ longitudinal excited states:
    \begin{equation}\label{long_exc}
        |\mathbf k,\mathbf q,\sigma\sigma\rangle=P_{G}\gamma_{\mathbf{k}\sigma+}^\dagger \gamma_{\mathbf{k}-\mathbf{q}\sigma-}|\psi_{\rm{MF}}\rangle
    \end{equation}
    At $\mathbf{q}=(0,0)$ and $\mathbf{q}=(\pi,\pi)$, we must complement this set with the ground state $P_{G}|\psi_{\rm{MF}}\rangle$ since they all belong to the same momentum in the magnetic Brillouin zone and thus might overlap. The numerical projection of the Heisenberg Hamiltonain on this set of states is performed in exactly the same way as in the transverse case. Note that calculating the amplitudes $\langle\beta|\mathbf{k}',\mathbf{q},\sigma\sigma\rangle$ now involves updates of the Slater determinant by changing for instance two rows and one column simultaneously, which can be done by generalizing the determinant update formula to arbitrary rank-$n$ change of rows and columns.

    To calculate $\mathcal{S}^{zz}(\mathbf{q},\omega)$, we derive the coefficients $\tilde\phi_{\mathbf{k}\mathbf{q}\sigma}^0$ such that
    \begin{equation}
        S_{\mathbf{q}}^z|\psi_{\rm MF}\rangle=\sum_{\mathbf{k,\sigma}}\tilde\phi^0_{\mathbf{k}\mathbf{q}\sigma}
        |\mathbf{k},\mathbf{q},\sigma\sigma\rangle 
        +\delta_{\mathbf q\mathbf \Pi}\sum_{\mathbf k}\frac{H_{\rm N}}{\omega_{\mathbf k}}
        \left|\psi_{\rm MF}\right\rangle\, .
    \end{equation}
    An explicit calculation gives
    \begin{equation}
        \tilde\phi^0_{\mathbf{k}\mathbf{q}\sigma}=
        \sigma\left(
        u_{\mathbf{k}\sigma+}^* u_{\mathbf{k}-\mathbf{q}\sigma-}
        +v_{\mathbf{k}\sigma+}^* v_{\mathbf{k}-\mathbf{q}\sigma-}
        \right)\, .
    \end{equation}

    As in the transverse case, the longitudinal dynamic structure factor is obtained by restricting the sum over the excited states to the set of projected longitudinal particle-hole states (Eq.~S\ref{long_exc}) giving
    \begin{equation}
        \mathcal{S}^{zz}(\mathbf q,\omega)=\sum_{\mathbf k}\left|\langle\mathbf q, n, 0|S_{\mathbf q}^z|\mbox{GS}\rangle\right|^2\delta(\omega-E_{\mathbf q n}^0+E_{\rm GS})
    \end{equation}
    where $|\mathbf q,n,0\rangle$ is the longitudinal eigenstate corresponding the eigenvalue $E_{\mathbf q n}^0$ obtained by diagonalizing the Heisenberg spin Hamiltonian (Eq. \ref{heis_spin}) onto the non-orthonormal states $|\mathbf q,\mathbf k,\sigma\sigma\rangle$ (Eq.~S\ref{long_exc}).

    The normalization of the longitudinal structure factor is performed in the same way as in the transverse case from the sum rule
    \begin{equation}
        \int d\omega \mathcal{S}^{zz}(\mathbf{q},\omega)=\langle S^z_{-\mathbf{q}}S^z_{\mathbf{q}} \rangle\, .
    \end{equation}

    In Fig.\ S\ref{fig7}, we show the longitudinal dynamic structure factor calculated in the $|\mbox{SF{+}N}\rangle$ state. Note that for the $|\mbox{SF}\rangle$ state, the longitudinal structure factor coincides with the transverse one.

    \subsection{Neel-field effect on fractional quasiparticle deconfinement}

    \begin{figure}[t!]
        \centering
        \includegraphics[width=\linewidth]{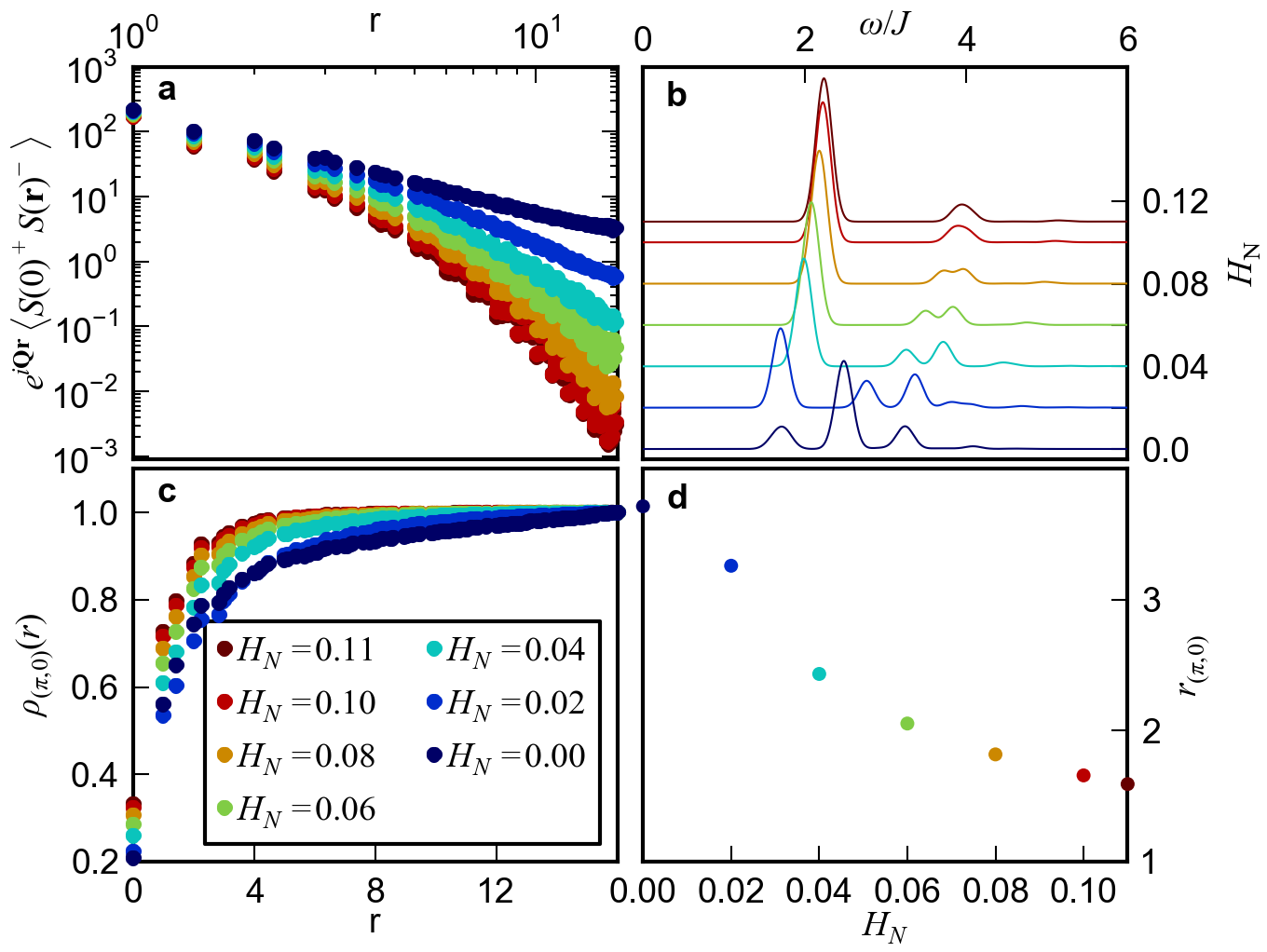}
        \caption{\label{fig10} Dependence of the ground-state spin correlations and excitations at $(\pi,0)$ in the $|\mbox{SF{+}N}\rangle$ state on the variational N\'{e}el field $H_{\rm N}$. The system size is $32\times 32$ and the value of the staggered flux is kept at the optimal level for the $|\mbox{SF{+}N}\rangle$ state ($\theta_0=\pi/10$). ({\bf a}) The equal-time transverse spin correlation function in real space in the log-log scale. ({\bf b}) The transverse dynamic structure factor. ({\bf c}) The disk-integrated quasiparticle-pair separation distribution (defined in the same way as in the panels {\bf c} and {\bf d} of Fig.\ \ref{fig5}). ({\bf d}) The root-mean-square quasiparticle-pair separation (defined in the same way as in the panel {\bf e} of Fig.\ \ref{fig5}). Lines and marker colors corresponds to panel C legend.}
    \end{figure}

    In this section, we provide additional data on the variational excitations and ground-state properties as a function of the variational field $H_{\rm N}$. As $H_{\rm N}$ interpolates between the $|\mbox{SF}\rangle$ state ($\theta_0=0.1\pi$, $H_{\rm N}=0$) and the $|\mbox{SF{+}N}\rangle$ state (the optimal values $\theta_0=0.1\pi$ and $H_{\rm N}=0.11$), the transverse spin fluctuations change from algebraic to exponential (Fig.\ S\ref{fig10}{\bf a}), the continuum in the variational spectrum is pushed to higher energies leaving behind a magnon peak (Fig.\ S\ref{fig10}{\bf b}), and the spatial extent of the $S=1/2$ quasiparticles pairs decreases (Figs.\ S\ref{fig10}{\bf c} and S\ref{fig10}{\bf d}). This behavior is consistent with our conjecture that the long-distance transverse spin-fluctuations play a crucial role for the possibility of fractional quasiparticle deconfinement. 


    \end{document}